\documentclass[onecolumn]{IEEEtran}

\usepackage{amsmath,stmaryrd,acronym,amssymb,amsthm,dsfont,graphicx}
\graphicspath{{graphics/}}

\acrodef{AEP}{Asymptotic Equipartition Property}
\acrodef{AoA}{Angle of Arrival}
\acrodef{AWGN}{Additive White Gaussian Noise}
\acrodef{BER}{Bit-Error-Rate}
\acrodef{BEC}{Binary Erasure Channel}
\acrodef{BPSK}{Binary Phase-Shift Keying}
\acrodef{BSC}{Binary Symmetric Channel}
\acrodef{CDF}[CDF]{Cumulative Distribution Function}
\acrodef{CLT}[CLT]{Central Limit Theorem}
\acrodef{CSI}[CSI]{Channel State Information}
\acrodef{DMC}[DMC]{Discrete Memoryless Channel}
\acrodef{DMS}[DMS]{Discrete Memoryless Source}
\acrodef{iid}[i.i.d.]{independent and identically distributed}
\acrodef{lhs}[l.h.s.]{left-hand-side}
\acrodef{rhs}[r.h.s.]{right-hand-side}
\acrodef{LPD}[LPD]{Low Probability of Detection}
\acrodef{LDPC}[LDPC]{Low-Density Parity-Check}
\acrodef{MAC}[MAC]{multiple-access channel}
\acrodef{MIMO}[MIMO]{Multiple-Input Multiple-Output}
\acrodef{MISO}{Multiple-Input Single-Output}
\acrodef{PDF}[PDF]{Probability Distribution Function}
\acrodef{PMF}[PMF]{Probability Mass Function}
\acrodef{PPM}[PPM]{Pulse Position Modulation}
\acrodef{PSD}{Power Spectral Density}
\acrodef{QPSK}{Quadrature Phase-Shift Keying}
\acrodef{SIMO}{Single-Input Multiple-Output}
\acrodef{SNR}{Signal-to-Noise Ratio}
\acrodef{wrt}[w.r.t.]{with respect to}
\acrodef{WSS}{Wide Sense Stationary}

\DeclareMathAlphabet{\eurm}{U}{eur}{m}{n}
\DeclareMathAlphabet{\mathbsf}{OT1}{cmss}{bx}{n}
\DeclareMathAlphabet{\mathssf}{OT1}{cmss}{m}{sl}
\DeclareMathAlphabet{\mathcsf}{OT1}{cmss}{sbc}{n}

\DeclareSymbolFont{bsfletters}{OT1}{cmss}{bx}{n}  
\DeclareSymbolFont{ssfletters}{OT1}{cmss}{m}{n}
\DeclareMathSymbol{\bsfGamma}{0}{bsfletters}{'000}
\DeclareMathSymbol{\ssfGamma}{0}{ssfletters}{'000}
\DeclareMathSymbol{\bsfDelta}{0}{bsfletters}{'001}
\DeclareMathSymbol{\ssfDelta}{0}{ssfletters}{'001}
\DeclareMathSymbol{\bsfTheta}{0}{bsfletters}{'002}
\DeclareMathSymbol{\ssfTheta}{0}{ssfletters}{'002}
\DeclareMathSymbol{\bsfLambda}{0}{bsfletters}{'003}
\DeclareMathSymbol{\ssfLambda}{0}{ssfletters}{'003}
\DeclareMathSymbol{\bsfXi}{0}{bsfletters}{'004}
\DeclareMathSymbol{\ssfXi}{0}{ssfletters}{'004}
\DeclareMathSymbol{\bsfPi}{0}{bsfletters}{'005}
\DeclareMathSymbol{\ssfPi}{0}{ssfletters}{'005}
\DeclareMathSymbol{\bsfSigma}{0}{bsfletters}{'006}
\DeclareMathSymbol{\ssfSigma}{0}{ssfletters}{'006}
\DeclareMathSymbol{\bsfUpsilon}{0}{bsfletters}{'007}
\DeclareMathSymbol{\ssfUpsilon}{0}{ssfletters}{'007}
\DeclareMathSymbol{\bsfPhi}{0}{bsfletters}{'010}
\DeclareMathSymbol{\ssfPhi}{0}{ssfletters}{'010}
\DeclareMathSymbol{\bsfPsi}{0}{bsfletters}{'011}
\DeclareMathSymbol{\ssfPsi}{0}{ssfletters}{'011}
\DeclareMathSymbol{\bsfOmega}{0}{bsfletters}{'012}
\DeclareMathSymbol{\ssfOmega}{0}{ssfletters}{'012}

\newcommand{\calC}{{\mathcal{C}}}
\newcommand{\calD}{{\mathcal{D}}}

\newcommand{\calI}{{\mathcal{I}}}

\newcommand{\calX}{{\mathcal{X}}}
\newcommand{\calY}{{\mathcal{Y}}}

\newcommand{\calZ}{{\mathcal{Z}}}

\newcommand{\E}[2][]{{\mathbb{E}_{#1}}{\left(#2\right)}}       
\renewcommand{\P}[2][]{{\mathbb{P}_{#1}}{\left(#2\right)}}
\newcommand{\Var}[1]{{\text{\textnormal{Var}}{\left(#1\right)}}}       
\newcommand{\D}[2]{{{\mathbb{D}}\!\left({#1\Vert#2}\right)}}
\newcommand{\avgD}[2]{{{\mathbb{D}}\!\left({#1\Vert#2}\right)}}
\newcommand{\V}[1]{{{\mathbb{V}}\!\left(#1\right)}}

\newcommand{\Hb}[1]{{\mathbb{H}_b}\left(#1\right)}

\newcommand{\wt}[1]{\ensuremath{\textnormal{wt}(#1)}}

\newcommand{\card}[1]{\ensuremath{\left|{#1}\right|}}           
\newcommand{\abs}[1]{\ensuremath{\left|#1\right|}}              
\newcommand{\eqdef}{\ensuremath{\triangleq}}                    
\newcommand{\intseq}[2]{\ensuremath{\llbracket{#1},{#2}\rrbracket}}  
\newcommand{\indic}[1]{\ensuremath{\mathds{1}\!\left\{#1\right\}}}

\renewcommand{\leq}{\leqslant}
\renewcommand{\geq}{\geqslant}

\newcommand{\proddist}{%
  \mathchoice{\raisebox{1pt}{$\displaystyle\otimes$}}
             {\raisebox{1pt}{$\otimes$}}
             {\raisebox{0.5pt}{\scalebox{0.7}{$\scriptstyle\otimes$}}}
             {\raisebox{0.4pt}{\scalebox{0.6}{$\scriptscriptstyle\otimes$}}}}
\newcommand{\pn}{{\proddist n}}

\usepackage[textwidth=0.68in,textsize=footnotesize,disable]{todonotes}
\presetkeys{todonotes}{color=redArcom, linecolor=redArcom,bordercolor=redArcom}{}

\newtheorem{theorem}{Theorem}
\newtheorem{remark}{Remark}
\newtheorem{definition}{Definition}
\newtheorem{lemma}{Lemma}

\acrodef{ROC}[ROC]{Receiver Operation Characteristic}
\acrodef{PPM}[PPM]{Pulse-Position Modulation}
\newcommand{\modified}[1]{{\color{black} #1}}
\newcommand{\typo}[1]{{\color{black} #1}}
\newcommand{\pr}[1]{{\left(#1\right)}}

\newcommand{\ppmX}[2]{P_{\mathbf{X}, \textnormal{PPM}}^{#1, #2}}
\newcommand{\ppmY}[2]{P_{\mathbf{Y}, \textnormal{PPM}}^{#1, #2}}
\newcommand{\ppmZ}[2]{P_{\mathbf{Z}, \textnormal{PPM}}^{#1, #2}}

\newcommand{\editt}[1]{{\color{black}#1}}
\begin{document}

\title{\modified{\editt{First and Second Order Asymptotics in Covert Communication}}}
\author{Mehrdad Tahmasbi and Matthieu R. Bloch \thanks{This work was presented in part at the 2016 IEEE International Symposium on Information Theory~\cite{Bloch2016}. This work was supported by NSF under award TWC 1527387.}}
\maketitle

\begin{abstract}
\editt{We study {the first- and } second-order asymptotics of covert communication over binary-input \acp{DMC} for three different covertness metrics and under maximum probability of error constraint. When covertness is measured in terms of the relative entropy between the channel output distributions induced with and without communication, we characterize the exact first- and second-order asymptotics of the number of bits that can be reliably transmitted with a maximum probability of error less than $\epsilon$ and a relative entropy less than $\delta$. When covertness is measured in terms of the variational distance between the channel output distributions or in terms of the probability of missed detection for fixed probability of false alarm, we establish the exact first-order asymptotics and bound the second-order asymptotics. {\ac{PPM} achieves the optimal first-order asymptotics for all three metrics, as well as the optimal second-order asymptotics for relative entropy.} The main conceptual contribution of this paper is to clarify how the choice of a covertness metric impacts the information-theoretic limits of covert communications. The main technical contribution underlying our results is a detailed expurgation argument to show the existence of a code satisfying the reliability and covertness criteria.}
\end{abstract}

\section{Introduction}
\label{sec:introduction}
While most information-theoretic security works to date have revolved around the issues of confidentiality and authentication~\cite{Bloch2011,InfoTheoreticSec,Jorswieck2015}, the growing concern around mass communication surveillance programs has reignited interest in investigating the covertness of communications, also known as \ac{LPD}. In \ac{LPD} problems, the objective is to hide the presence of communication and not necessarily to prevent information leakage about the messages transmitted. Following the analysis of \ac{LPD} with space-time codes~\cite{Hero2003}, recent works have \editt{investigated} the information-theoretic limits of covert communications over noisy channel~\cite{Bash2013,Korzhik2005}. In particular, building upon concepts from steganography~\cite{Cachin2004}, \editt{the study in~\cite{Bash2013} shows the existence of a ``square-root law'' for covert communication}, which essentially states that no more than $O(\sqrt{n})$ bits can be communicated covertly over $n$ channel uses of a memoryless channel. 

The square-root law of covert communication has been refined in several follow-up works~\cite{Che2013}; in particular, the information-theoretic limits are now known for classical discrete and Gaussian memoryless channels~\cite{Bloch2016a,Wang2016b}, classical-quantum channels~\cite{Wang2016c,Sheikholeslami2016}, and multiple-access channels~\cite{Arumugam2016}, when covertness is measured in terms of the relative entropy between the channel output distributions induced with and without communication. Note that the choice of relative entropy as a metric for covertness is guided in part by the natural connection between relative entropy and information-theoretic metrics, such as entropy and mutual information, which have been largely explored in the context of information-theoretic security~\cite{Hou2014}.

The contribution of the present paper is twofold. First, as an attempt to develop operational characterizations of covertness, we study the information-theoretic limits of covert communication for alternative metrics, including variational distance, and probability of missed detection. Second, motivated by the likely time-limited nature of covert communications, we make a first step towards a \editt{finite-length} analysis and extend the first-order analysis of information-theoretic limits to second-order asymptotics. The specific results developed in the present paper focus on binary-input \acp{DMC} to \editt{yield simple closed-form} expressions and are the following.
\begin{itemize}
\item We characterize the exact second-order asymptotics of the maximum number of reliable and covert bits that can be transmitted with \editt{maximum}  probability of error $\epsilon$ and \emph{relative entropy} $\delta$ between channel output distributions with and without communication (Theorem~\ref{th:main_result_d}); this corrects an unfortunate error in the conference version~\cite{Bloch2016}, in which we claimed erroneous second-order asymptotics for arbitrary codes. 
\item \editt{We characterize the exact first-order and bound the second-order asymptotics of the maximum number of reliable and covert bits that can be transmitted with maximum  probability of error $\epsilon$ and \emph{variational distance} $\delta$ between channel output distributions with and without communication (Theorem~\ref{th:main_result_tv}).} 
\item \editt{Finally, we characterize the exact first-order and bound the second-order asymptotics of the maximum number of reliable and covert bits that can be transmitted with maximum  probability of error $\epsilon$ and probability of missed detection $1-\alpha-\delta$ when $\alpha$ is the adversary's probability of false alarm (Theorem~\ref{th:main_result_beta}).} 
\end{itemize}
\editt{All our achievability results are established using \acf{PPM}, which optimality was previously only established for the first-order asymptotics with relative entropy~\cite{Bloch2016b}. The operational relevance of codes used in conjunction with \ac{PPM}, which may be viewed as a highly structured subset of constant composition codes, is justified by recent work towards practical code design~\cite{Kadampot2018}, in which \ac{PPM} plays a crucial role. We also emphasize from the outset that the focus on maximal probability of error is essential to our analysis. As discussed in Remark~\ref{rem:average-prob-error} and Appendix~\ref{sec:effect-aver-prob}, there is no strong converse for average probability of error without additional constraint; hence, considering the \emph{maximum} probability of error as the reliability metric is reasonable for second-order analysis.}

The second-order asymptotics obtained with a relative entropy metric for covertness are what could have been expected by extrapolating the results of channel coding in the finite length regime~\cite{Polyanskiy2010} to the first-order asymptotics of covert communication~\cite{Bloch2016a,Wang2016b}; however, the proof requires specific techniques beyond those used to study the first-order asymptotics of covert communications \typo{and second-order asymptotics of classical communication}. First, as already mentioned, the achievability proof relies on \ac{PPM} codes~\cite{Bloch2016b} instead of \ac{iid} random codes. Second, \typo{unlike the analysis of the second-order asymptotics of reliable and secure communications~\cite{Polyanskiy2010,Yang2016}, we have to deal here with parameters capturing reliability and covertness, with the latter appearing in the first-order asymptotics; more specifically,} the optimal coding scheme identified in~\cite{Bloch2016a} exploits a code with a bin structure, in which each bin forms a reliability code for the legitimate channel indexed by the secret key while the overall code forms a resolvability code for the adversary's channel. To prove \editt{the existence} of a code with the desired characteristics, \editt{we expurgate a random code after resorting to concentration of measure inequalities such as McDiarmid's Inequality and carefully analyzing the probability of error}. We point out that our current results only identify the second-order asymptotics for the number of transmitted message bits and do not characterize the second-order asymptotics for the number of key bits.

The remainder of the paper is organized as follows. In Section~\ref{sec:main-result}, we formally introduce the model of covert communication and state our main results. In Section~\ref{sec:covert-comm-results}, we develop a series of \editt{metric-independent} results that form the basis of our analysis of covert communication. In Section~\ref{sec:covert-comm-with-kl}, Section~\ref{sec:secon-order-covert-tv}, and Section~\ref{sec:covert-comm-missed-detection}, we exploit the results of Section~\ref{sec:covert-comm-results} to study a relative entropy metric, variational distance metric, and probability of missed-detection metric for covertness, respectively. In Section~\ref{sec:conclusion}, we conclude the paper with a discussion of possible further extensions and improvements.

\section{Model and main results}
\label{sec:main-result}

\subsection{Notation}
\label{sec:notation}

\todo[inline]{We need to define all important quantities, say D, V, $\beta$, base of log, notation for random variables, and distributions.}

Throughout the paper, $\log$ and $\exp$ should be understood in base $e$. Moreover, random variables are denoted with upper case letters, e.g. $X$, while their \editt{realizations} are denoted with lower case letters, e.g., $x$. The distribution of a random variable such as $X$ is denoted by $P_X$. Calligraphic letters are used for sets, e.g., $\mathcal{X}$, and boldface fonts are used for vectors  e.g., $\textbf{x}$. For two integers $a$ and $b$, if $a\leq b$, we define $\intseq{a}{b}\eqdef \{a, a+1, \cdots, b-1, b\}$; otherwise $\intseq{a}{b} \eqdef \emptyset$. For any distribution $P$ over $\mathcal{X}$, $P^\pn$ denotes the product distribution over $\mathcal{X}^n$, i.e., $P^\pn(\textbf{x}) \eqdef \prod_{i=1}^n P(x_i)$. For two distributions $P$ and $Q$ over the same set $\mathcal{X}$, we define
\begin{align}
\D{P}{Q} &\eqdef \sum_x P(x) \log \frac{P(x)}{Q(x)} \quad\text{(relative entropy)},\\
\V{P, Q} &\eqdef \frac{1}{2}\sum_x |P(x) - Q(x)| \quad\text{(variational distance)},\\
\chi_2(P||Q) &\eqdef \sum_{x} \frac{(P(x)-Q(x))^2}{Q(x)}\quad\text{(chi-squared distance)},\\
\beta_{\alpha}(P, Q) &\eqdef \inf_{T \subset \mathcal{X}: P(\modified{\calX\setminus}T) \leq \alpha} Q(T) \quad\text{(optimal probability of missed detection)}.
\end{align}
The notation $P\ll Q$  means that $P$ is absolutely continuous \ac{wrt} $Q$, i.e., if $Q(x)=0$ for some $x\in \mathcal{X}$ then $P(x) = 0$.  A \ac{DMC} $(\mathcal{X}, W_{Y|X}, \mathcal{Y})$ consists of a finite input alphabet $\mathcal{X}$, a finite output alphabet $\mathcal{Y}$, and a transition probability $W_{Y|X}$ \editt{such that} $W_{Y|X}(y|x)$ indicates the probability of obtaining $y$ at the output given that $x$ is transmitted at the input. For a \ac{DMC} $(\mathcal{X}, W_{Y|X}, \mathcal{Y})$ and two distributions $P_X$ and $Q_Y$ on $\mathcal{X}$ and $\mathcal{Y}$, we define
\begin{align}
P_Y(y) &\eqdef \sum_x P_X(x) W_{Y|X}(y|x),\\
I(P_X, W_{Y|X}) &\eqdef \sum_{x, y}  P_X(x) W_{Y|X}(y|x) \log \frac{W_{Y|X}(y|x)}{P_Y(y)},\label{eq:I_def}
\end{align}
 and for any $\gamma$,
\begin{align}
F_{XY|Q_Y}(\gamma) &\eqdef \P[W_{Y|X}P_X]{\log \frac{W_{Y|X}(Y|X)}{Q_Y(Y) }\leq \gamma} = \sum_{x, y} P_X(x) W_{Y|X} (y|x) \indic{\log \frac{W_{Y|X}(y|x)}{Q_Y(y) }\leq \gamma},\\
\editt{F_{XY|Q_YX=x}(\gamma) }&\eqdef \P[W_{Y|X}P_X]{\log \frac{W_{Y|X}(Y|X)}{Q_Y(Y) }\leq \gamma\Bigg| X=x} = \sum_{y}  W_{Y|X} (y|x) \indic{\log \frac{W_{Y|X}(y|x)}{Q_Y(y) }\leq \gamma},\\
F_{XY}(\gamma) &\eqdef F_{XY|P_Y}(\gamma),\\
\overline{F}_{XY}(\gamma) &\eqdef 1-F_{XY}(\gamma).
\end{align}
Moreover,  given codewords $\{x_i\}_{i=1}^M \in \calX^M$ and a uniform random variable $W\in\intseq{1}{M}$, $\widehat{P}_{WXY}$ denotes the joint distribution induced on $(W, X, Y)$, i.e.,
\begin{align}
\widehat{P}_{WXY}(w, x, y) \eqdef \frac{1}{M}\indic{x_w = x} W_{Y|X}(y|x).
\end{align}
For any discrete random variable $A$, let $\mu_A \eqdef \min_{a: \P{A=a} > 0} \P{A=a}$. \modified{For a real number $x$, we also define the $Q$-function $Q(x) \eqdef \int_{x}^\infty \frac{1}{\sqrt{2\pi}} e^{-\frac{x^2}{2}} \text{d}x$ and $[x]^+ \eqdef \max(0, x)$. Moreover, for sequences $\{a_n\}$, $\{b_n\}$, and $\{c_n\}$, we have $a_n = b_n + O(c_n)$ if and only if there exists a constant $K$ independent of $n$ such that $|a_n - b_n| \leq K c_n$ for all $n$. }
Finally, for $\textbf{x}\in\mathcal\{0,1\}^n$, \editt{$\wt{\textbf{x}}\eqdef \card{\{i\in\intseq{1}{n}:x_i=1\}}$ is the weight of $\mathbf{x}$}.

\subsection{Model and main results}
\label{sec:model-main-results}

\begin{figure}
  \centering
  \includegraphics[width=0.6\linewidth]{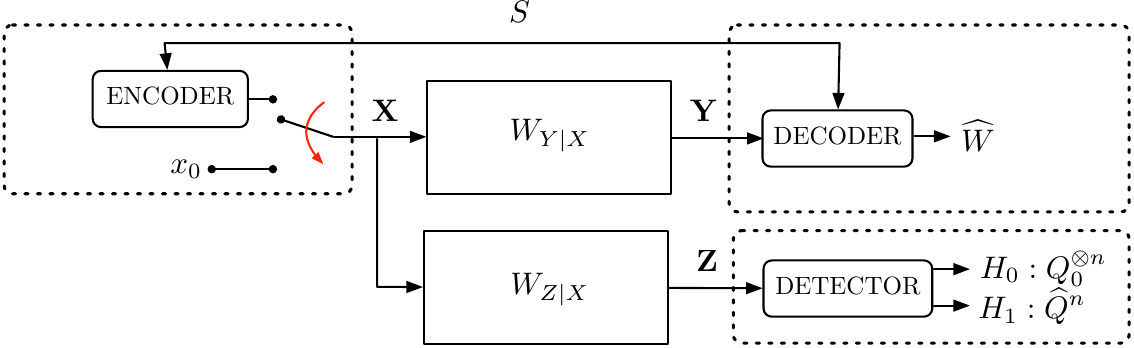}
  \caption{Model of covert communications over noisy channels}
  \label{fig:model}
\end{figure}

We consider the situation illustrated in Fig.~\ref{fig:model}, in which a legitimate transmitter communicates with a legitimate receiver over a \ac{DMC} $(\calX,W_{Y|X},\calY)$ in the presence of an adversary who observes communication through another \ac{DMC} $(\calX,W_{Z|X},\calZ)$. \editt{There is no loss in generality in ignoring the joint channel $W_{YZ|X}$, as we shall see that results only depend on the marginal channels $W_{Y|X}$ and $W_{Z|X}$.}  Furthermore,  \editt{to obtain simple closed-form expressions}, we assume in the sequel that $\calX=\{0,1\}$, where $0$ is an ``innocent symbol'' corresponding to the expected input of the channel if no communication were taking place. The extension beyond binary-inputs is briefly outlined in Section~\ref{sec:conclusion}. We also denote the output distribution induced by each input symbol by
\begin{align}
P_0\eqdef W_{Y|X=0},&\quad   P_1\eqdef W_{Y|X=1},\displaybreak[0]\\
Q_0\eqdef W_{Z|X=0},&\quad Q_1\eqdef W_{Z|X=1},
\end{align}
where it is assumed that $Q_1 \ll Q_0$, $P_1 \ll P_0$ and $Q_1\neq Q_0$. These assumptions are \typo{necessary for our results to hold. In fact, without $Q_1 \ll Q_0$, covert communication is impossible as some input symbols detect the use of the non-innocent symbol with probability one. At the other extreme,  if $Q_1 = Q_0$ then no detector can distinguish the use of innocent and non-innocent symbols so that the problem reduces to classical communication. Finally, without $P_1\ll P_0$, Alice and Bob have an unfair advantage that allows them to exchange $\omega(\sqrt{n})$ bits covertly \cite[Theorem 7]{Bloch2016a}.}
The objective is for the transmitter to communicate a uniformly distributed message $W\in\intseq{1}{M}$ with a small probability of error while ensuring low probability of detection from the adversary. The communication may be assisted by a uniformly distributed shared secret key $S\in\intseq{1}{K}$. A formal description of a code is as follows.
\begin{definition}
\label{def:covert_code}
 A code of blocklength $n$ consists of an encoder/decoder pair
\begin{align}
f:\intseq{1}{K}\times\intseq{1}{M}\rightarrow\calX^n\text{ and }\phi:\calY^n\times\intseq{1}{K}\rightarrow\intseq{1}{M},
\end{align}
which induces the distribution $\widehat{P}_{\mathbf{Z}}$ at the adversary's channel output, defined \editt{for $\mathbf{z}\in\calZ^n$ as}
\begin{align}
\widehat{P}_{\mathbf{Z}}(\textbf{z})\eqdef \sum_{s=1}^K\sum_{w=1}^M \frac{1}{MK}W_{Z|X}^\pn(\mathbf{z}|f(s,w)).
\end{align}
A code is $\epsilon$-reliable, if the pair $(f,\phi)$ is such that
\begin{align}
\editt{P_{\textnormal{err}}^*\eqdef\max_{s\in\intseq{1}{K}, w\in\intseq{1}{M}}\P{\phi(\mathbf{Y},s)\neq W|S=s, W=w}\leq \epsilon.}\label{eq:prob_error}
\end{align}
 Moreover, an $\epsilon$-reliable code defined by $(f,\phi)$ is:
\begin{itemize}
\item an $(M,K,n,\epsilon,\delta)_D$ code, if it satisfies
\begin{align}
\D{\widehat{P}_{\mathbf{Z}}}{Q_0^\pn}\leq \delta;\label{eq:covert_with_D}
\end{align}

\item an $(M,K,n,\epsilon,\delta)_V$ code, if it satisfies
\begin{align}
\V{\widehat{P}_{\mathbf{Z}},Q_0^\pn}\leq \delta;\label{eq:covert_with_V}
\end{align}

\item an $(M,K,n,\epsilon,\delta, \alpha)_\beta$ code, if if it satisfies
\begin{align}
\beta_\alpha(Q_0^{\pn}, \widehat{P}_{\mathbf{Z}}) \geq 1 - \alpha - \delta.\label{eq:covert_with_beta}
\end{align}
\end{itemize}
The maximum number of messages that can be transmitted by an $(M,K,n,\epsilon,\delta)_D$, $(M,K,n,\epsilon,\delta)_V$, and $(M,K,n,\epsilon,\delta, \alpha)_\beta$ code is denoted by $M^*_D(n, \epsilon, \delta)$, $M^*_V(n, \epsilon, \delta)$, and $M^*_\beta(n, \epsilon, \delta, \alpha)$, respectively. 
\end{definition}

\begin{remark}
  \label{rem:average-prob-error}
  The definition of the probability of error in~(\ref{eq:prob_error}) differs from previous studies~\cite{Wang2016b, Bloch2016a}; we ask that the \editt{maximum} probability of error be small for \emph{any} choice of the key $S$ and message $W$. This more stringent condition  captures a pragmatic requirement \editt{and is critical to our converse argument}. \editt{As shown in Appendix A, analyzing the average probability of error is a different and (we believe) substantially more difficult endeavor since there is no strong converse for the reliability parameter. This \editt{intriguing} behavior appears because one can slightly \editt{increase} the \editt{average probability of error} without any effect on covertness by simply adding all-zero codewords to the codebook. This of course leads to what one would consider as ``bad codes,'' which we avoid by focusing on the maximal probability of error.}
\end{remark}

\editt{
\begin{remark}
\textcolor{black}{The three covertness metrics in~(\ref{eq:covert_with_D})-(\ref{eq:covert_with_beta}) have \editt{slightly} different operational meanings. When enforcing \editt{that the probability of missed detection satisfy} $\beta_\alpha(Q_0^{\pn}, \widehat{P}_{\mathbf{Z}}) \geq 1 - \alpha - \delta$ for a fixed probability of false alarm $\alpha$, one implicitly assumes that the adversary \editt{optimizes its detector at a specific point of the \ac{ROC} curve with known probability of false alarm}. In contrast, when enforcing $\V{\smash{\widehat{P}_{\mathbf{Z}}},Q_0^\pn}\leq \delta$, since $\alpha+\beta_\alpha(Q_0^{\pn}, \widehat{P}_{\mathbf{Z}}) \geq 1-\V{\smash{\widehat{P}_{\mathbf{Z}}},Q_0^\pn}$ for \emph{any} probability of false alarm $\alpha$, one essentially wishes to enforce covertness irrespective of the exact operating point on the adversary's \ac{ROC} curve. Finally, since \typo{$\V{\smash{\widehat{P}_{\mathbf{Z}}},Q_0^\pn}^2\leq\frac{1}{2}{\D{\smash{\widehat{P}_{\mathbf{Z}}}}{Q_0^\pn}}$} \todo{Check that we are using the right base and definition for v} by Pinsker's Inequality, the constraint $\D{\smash{\widehat{P}_{\mathbf{Z}}}}{Q_0^\pn}\leq \delta$ is more stringent than when using variational distance, but \editt{only provides a loose proxy for constraining the \ac{ROC} curve since Pinsker's inequality is not tight.}.\footnote{See also the discussion in~\cite[Appendix~A]{Bloch2016a}.} The relations between the \editt{covertness metrics} immediately lead to the following ordering of the maximum number of covert bits.
\modified{
\begin{align}
  \label{eq:ordering}
  M^*_D(n, \epsilon, \delta) &\leq   M^*_V(n, \epsilon, \sqrt{\frac{\delta}{2}}) \leq \min_{\alpha\in[0,1]}M^*_\beta(n, \epsilon, \sqrt{\frac{\delta}{2}}, \alpha).
\end{align}
}}
\end{remark}
}

Our main results in Theorem~\ref{th:main_result_d}, Theorem~\ref{th:main_result_tv}, and Theorem~\ref{th:main_result_beta} characterize the maximum number of reliable and covert bits defined above as a function of the channel characteristics and the blocklength $n$. In all cases, the first and second terms behave as $\Theta\left(\smash{n^{\frac{1}{2}}}\right)$ and $\Theta\left(\smash{n^{\frac{1}{4}}}\right)$, respectively, as expected; nevertheless, the constant behind $\Theta(\cdot)$ is metric-specific.
\begin{theorem}
\label{th:main_result_d}
For $\epsilon \in ]0, 1[$ and $\delta>0$, we have
\begin{align} 
\editt{\log M^\text{*}_D(n, \epsilon, \delta)} = \sqrt{\frac{2\delta}{\chi_2(Q_1\|Q_0)}}D_Pn^{\frac{1}{2}}-\sqrt{\sqrt{\frac{2\delta}{\chi_2(Q_1\|Q_0)}}V_P}Q^{-1}(\epsilon)n^{\frac{1}{4}} + O(\log n),
\end{align}
with
\begin{align}
\label{eq:DVP_definition}
D_P \eqdef \avgD{P_1}{P_0}, &\quad
V_P \eqdef  \Var{\left.\log\frac{P_1(Y)}{P_0(Y)}\right|P_1}.
\end{align}
This optimal \editt{number of message bits} is obtained with a first-order optimal number of key bits
\begin{align}
\log K = (1+\rho)\sqrt{\frac{2\delta}{\chi_2(Q_1\|Q_0)}}[D_Q-D_P]^+ n^{\frac{1}{2}},
\end{align}
where $\rho>0$ can be arbitrarily small and $D_Q=\avgD{Q_1}{Q_0}$. \editt{In addition,
\begin{align}
  \lim_{\epsilon\to 0}\lim_{n\rightarrow\infty}\frac{\log M_D^*(n,\epsilon,\delta)}{\sqrt{n}} =  \sqrt{\frac{2\delta}{\chi_2(Q_1\|Q_0)}}D_P.
\end{align}
}
\end{theorem}
\begin{theorem}
\label{th:main_result_tv}
For $\epsilon \in ]0, 1[$ and $\delta\in ]0,1[$, let $\Gamma \eqdef Q^{\editt{-1}}\pr{\frac{1-\delta}{2}}$. We have
\begin{align} 
\label{eq:v-out-bound}
\editt{\log M^\text{*}_V(n, \epsilon, \delta)} \leq \frac{2\Gamma D_P}{\sqrt{\chi_2(Q_1\|Q_0)}}n^{\frac{1}{2}}-\sqrt{\frac{2\Gamma V_P}{\sqrt{\chi_2(Q_1\|Q_0)}}}Q^{-1}(\epsilon)n^{\frac{1}{4}} + O(\log n),
\end{align}
and
\begin{align} 
\editt{\log M^\text{*}_V(n, \epsilon, \delta)} \geq \frac{2\Gamma D_P}{\sqrt{\chi_2(Q_1\|Q_0)}}n^{\frac{1}{2}}-\pr{\sqrt{\frac{2\Gamma V_P}{\sqrt{\chi_2(Q_1\|Q_0)}}}Q^{-1}(\epsilon) + \frac{2\sqrt{\pi} e^{\frac{\Gamma^2}{2}} D_P }{\sqrt{\Gamma}\chi_2(Q_1\|Q_0)^{\frac{1}{4}} }}n^{\frac{1}{4}} + O(\log n),
\end{align}
where $D_P$ and $V_P$ are as in \eqref{eq:DVP_definition}. This \editt{number of message bits} is obtained with a number of key bits\todo[inline]{Is the number of key bits optimal?}
\begin{align}
\log K = (1+\rho)\frac{2\Gamma D_P}{\sqrt{\chi_2(Q_1\|Q_0)}} [D_Q-D_P]^+ n^{\frac{1}{2}},
\end{align}
where $\rho>0$ can be arbitrarily small. \editt{In addition,
\begin{align}
  \lim_{\epsilon\to 0}\lim_{n\rightarrow\infty}\frac{\log M_V^*(n,\epsilon,\delta)}{\sqrt{n}}=\frac{2\Gamma D_P}{\sqrt{\chi_2(Q_1\|Q_0)}}.
\end{align}
}
\end{theorem}

\begin{theorem}
\label{th:main_result_beta}
For $\epsilon \in ]0, 1[$, $\alpha\in]0,1[$, and $\delta\in ]0, 1-\alpha[$, let $\Lambda \eqdef Q^{-1}(1 - \alpha - \delta)$ and $\Upsilon \eqdef Q^{-1}(\alpha)$. We have
\begin{align} 
\label{eq:b-out-bound}
\editt{\log M^\text{*}_\beta(n, \epsilon, \delta, \alpha)} \leq \frac{(\Lambda + \Upsilon) D_P}{\sqrt{\chi_2(Q_1\|Q_0)}}n^{\frac{1}{2}}-\sqrt{\frac{(\Lambda + \Upsilon)V_P}{\sqrt{\chi_2(Q_1\|Q_0)}}}Q^{-1}(\epsilon)n^{\frac{1}{4}} + O(\log n),
\end{align}
and
\begin{align} 
\editt{\log M^\text{*}_\beta(n, \epsilon, \delta, \alpha) }\geq \frac{(\Lambda + \Upsilon)D_P}{\sqrt{\chi_2(Q_1\|Q_0)}}n^{\frac{1}{2}}-\pr{\sqrt{\frac{(\Lambda + \Upsilon) V_P}{\sqrt{\chi_2(Q_1\|Q_0)}}}Q^{-1}(\epsilon) + \frac{\sqrt{2\pi}\pr{e^{\frac{\Gamma^2}{2}} + e^{\frac{\Upsilon^2}{2}}} D_P}{\sqrt{\Lambda +\Upsilon}\chi_2(Q_1\|Q_0)^{\frac{1}{4}}}} n^{\frac{1}{4}} + O(\log n).
\end{align}
This \editt{number of  message bits} is obtained with a number of key bits\todo[inline]{Is the number of key bits optimal?}
\begin{align}
\log K = (1+\rho)\frac{\pr{\Lambda + \Upsilon} [D_Q-D_P]^+}{\sqrt{\chi_2(Q_1\|Q_0)}} n^{\frac{1}{2}},
\end{align}
where $\rho>0$ can be arbitrarily small. \editt{In addition,
\begin{align}
  \lim_{\epsilon\to 0}\lim_{n\rightarrow\infty}\frac{\log M_\beta^*(n,\epsilon,\delta,\alpha)}{\sqrt{n}}  = \frac{(\Lambda + \Upsilon)D_P}{\sqrt{\chi_2(Q_1\|Q_0)}}.
\end{align}
}
\end{theorem}

\modified{
\editt{We point out that} the second order asymptotics in Theorem~\ref{th:main_result_tv} and Theorem~\ref{th:main_result_beta} are loose because the parameters $\Gamma$, $\Lambda$, and $\Upsilon$ may be very small; in particular, because of \editt{the inequalities}~\eqref{eq:ordering}, Theorem~\ref{th:main_result_d} \editt{is sometimes} a tighter lower bound for \editt{small} values of $n$. \editt{We conjecture that the upper-bounds in \eqref{eq:v-out-bound} and \eqref{eq:b-out-bound} can be achieved, although we could not establish it with our current proof techniques.}}

\begin{figure}[h]
  \centering
  \includegraphics[width=0.7\linewidth]{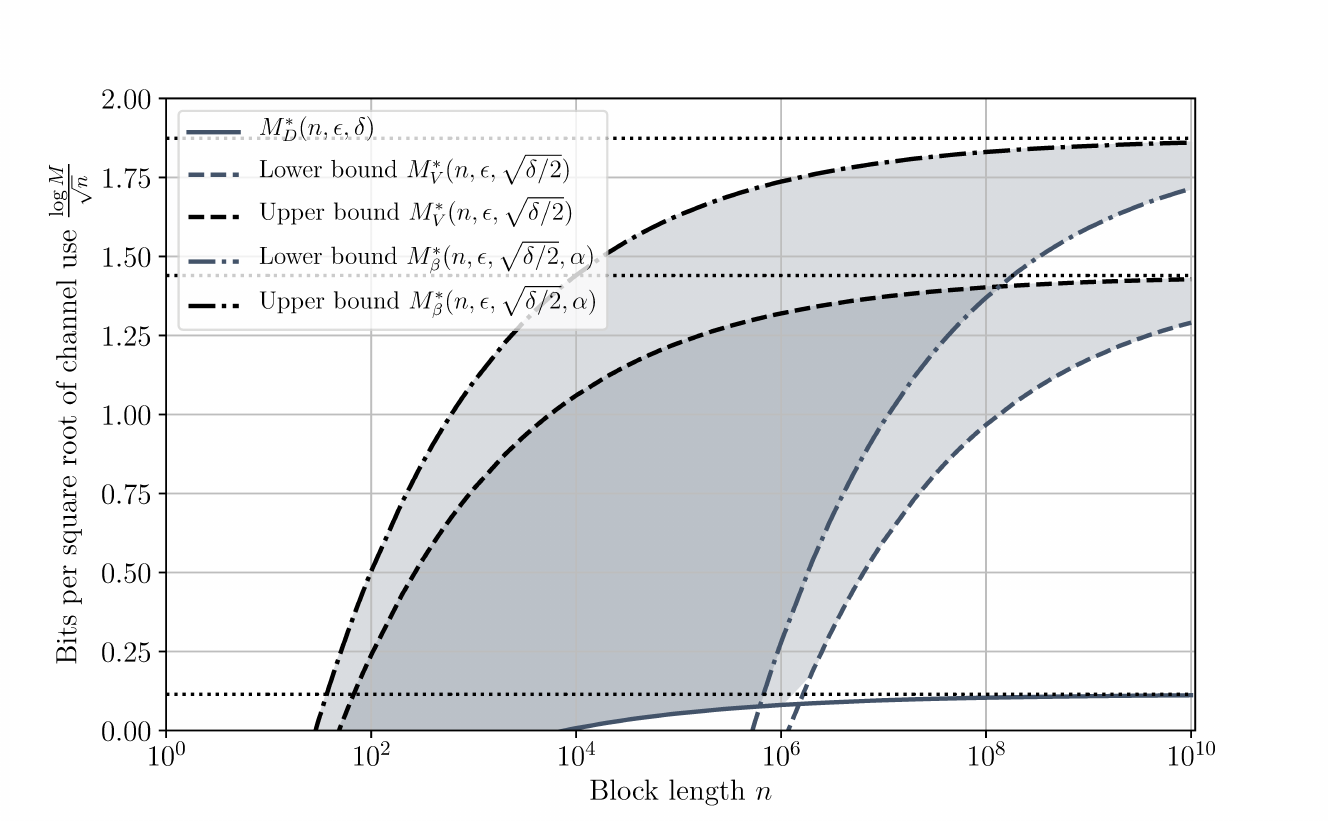}
  \caption{\editt{Second-order approximation of} maximum number of covert and reliable bits as a function of blocklength. Both \acp{DMC} are \acp{BSC} with cross-over probability $p_m=0.11$ and $p_w=0.45$, respectively, and $\epsilon=10^{-3}$, $\delta=10^{-2}$, and $\alpha=0.2$. \modified{The dotted horizontal lines indicate the optimal first-order asymptotics.}}
  \label{fig:illustration}
\end{figure}

\modified{We illustrate \editt{the results of the three theorems} with a simple numerical example. We consider the situation in which $(\calX, W_{Y|X}, \calY)$ and $(\calX, W_{Z|X}, \calZ)$ are \acp{BSC} with cross-over probability $p_m=0.11$ and $p_w=0.45$, respectively, and $\epsilon=10^{-3}$, $\delta=10^{-2}$, and $\alpha=0.2$. Since the convergence to the asymptotic limit is slow \editt{(on the order of $\Theta(n^{-\frac{1}{4}})$)}, we use a log scale for the blocklength.} As shown in Fig.~\ref{fig:illustration}, the choice of the covertness metric results in different number of bits, which of course raises the question of which number to settle on. We argue that $M^*_V(n, \epsilon, \delta)$ is the number to focus on since total variation distance satisfies two desirable properties: it is directly \editt{related} to the \editt{performance of the adversary's detector} through the inequality $\alpha+\beta \geq 1-\V{\smash{\widehat{P}_{\mathbf{Z}}},Q_0^\pn}$ and it does not presume any knowledge about the exact operating point on the adversary's \ac{ROC} curve.

Before we detail the achievability and converse proofs in the next sections, we provide here a high-level sketch of the proofs. Following~\cite{Bloch2016a}, the coding scheme in the achievability proof consists of $MK$ randomly generated codewords $\mathbf{x}_{sw}$ with $s\in\intseq{1}{K}$ and $w\in\intseq{1}{M}$. The code is designed such that the following \editt{two} properties hold:
\begin{itemize}
\item[(P1)] any ``large" enough subset of codebook  $\{\mathbf{x}_{sw}\}_{(s,w)\in\intseq{1}{K}\times\intseq{1}{M}}$ forms a resolvability code for the channel $(\calX,W_{Z|X},\calZ)$ approximating $Q_0^\pn$;
\item[(P2)] for every $s\in\intseq{1}{K}$, the sub-codebook $\{\mathbf{x}_{sw}\}_{w\in\intseq{1}{M}}$ forms a reliability code for the channel $(\calX,W_{Y|X},\calY)$ with a \editt{\emph{maximum}} probability of error $\epsilon$.
\end{itemize}
 The analysis does not follow from standard arguments for two reasons. First, \editt{we resort to \ac{PPM} codes~\cite{Bloch2016b} to  obtain the optimal second-order asymptotics for relative entropy}. Second, one cannot ensure (P1) and (P2) by merely expurgating some of the sub-codebooks, since this expurgation may change the output distribution on $\calZ^n$ induced by the code; we address this by carefully analyzing the probability of error and using concentration of measure results. The converse proof adapts arguments from~\cite{Polyanskiy2010} \editt{after showing how the covertness constraint \editt{leads to} an upper bound for the average codeword weight}.

\section{Covert communications with generic covertness quasi-metric}
\label{sec:covert-comm-results}

We now develop results for a generic covertness metric, which will be specialized to the three metrics highlighted in Definition~\ref{def:covert_code} in Section~\ref{sec:covert-comm-with-kl}, Section~\ref{sec:secon-order-covert-tv}, and Section~\ref{sec:covert-comm-missed-detection}. This organization allows us to \editt{separately handle} the part of the analysis that \editt{solely depends} on the code structure and not on the exact metric choice. Specifically, throughout this section, we consider an arbitrary \emph{quasi-metric} $d$ measuring the distance between distributions. The quasi-metric $d$ does not necessarily possess the standard characteristics of a metric; however, we assume that it is non-negative and satisfies a modified triangle inequality, i.e.,
\begin{itemize}
\item $\forall P, Q:\quad d(P, Q)\geq 0$
\item $\forall P, Q, R:\quad d(R, Q) \leq d(P, Q) + \D{R}{P}+\sqrt{\D{R}{P}}\max\left(1,\log \frac{1}{\min_{x:Q(x)>0} Q(x)}\right)$.
\end{itemize}
This modified version of the triangle inequality is intimately linked to the relative entropy, and is used mainly for convenience; this will avoid repetitions when we study different metrics and should not be given much operational significance. In brief, we will use the modified triangle inequality as follows: we will pick an appropriate distribution $P$ such that $d(R,Q)$ is essentially on the order of $d(P,Q)$ \editt{that} depends on $d$, while the remaining terms that depend on relative entropy will be made negligible. We define an $(M, K, n, \epsilon, \delta)_d$ covert code and  $M^*_d(n, \epsilon, \delta)$ for a covert communication channel $(\calX, W_{Y|X}, W_{Z|X}, \calY, \calZ)$ as in Definition~\ref{def:covert_code}.

\modified{Our proof for a generic covertness quasi-metric $d$ is organized as follows. In \editt{Section~\ref{sec:one-shot-achi}}, we start by developing one-shot results for random codes to bound the average probability of error (Lemma~\ref{lm:ch_coding}) and the approximation of output statistics as measured with the quasi-metric (Lemma~\ref{lm:ch_resolve_p} and Lemma~\ref{lm:ch_resolve_e}). The expurgation of the code requires some care because of the interplay between reliability and channel resolvability (Lemma~\ref{thm:one-shot-covert-cm}). In Section~\ref{sec:asympt-results-cover}, we specialize the one-shot results to obtain an achievability result for random codes with large-enough blocklength (Theorem~\ref{thm:general_achv}). In Section~\ref{sec:distributions}, we refine the statement of Theorem~\ref{thm:general_achv} for the specific random code ensemble consisting of \ac{PPM} codes. This requires the study of the specific quantities involved in the statement of Theorem~\ref{thm:general_achv} (Lemma~\ref{lm:F_XY_bound}, Lemma~\ref{lm:expected_ration_ppm}, Lemma~\ref{lm:F_XZ_bound}). We also pave the way for our final results by developing bounds between the output distribution induced by a \ac{PPM} ensemble and the innocent distribution for relative entropy, variational distance, and probability of missed-detection (Lemma~\ref{lm:div_ppm}). These bounds are fairly technical and rely on moment bounds (Lemma~\ref{lm:A-moments}). Finally, in Section~\ref{sec:generic-converse}, we develop a general converse (Theorem~\ref{thm:converse-general}).}

\subsection{One-shot achievability analysis}
\label{sec:one-shot-achi}
We develop one-shot random coding results for codes simultaneously ensuring channel reliability and channel resolvability. Given a \ac{DMC}$(\mathcal{X}, W_{Y|X}, \mathcal{Y})$ and a codebook with $M$ codewords $x_1, \cdots, x_M\in \mathcal{X}$,\footnote{\modified{Since we are deriving one-shot bounds, $\calX$ can be arbitrary in this section.}} our analysis of reliability is based on a threshold decoder~\cite{Polyanskiy2010} operating as follows. Given $\gamma>0$ and a distribution $Q_Y$ on $\mathcal{Y}$, and upon observing $y$, the decoder forms the estimate $\widehat{W}=w$ of the transmitted message $W$ if there exists a unique $w\in\intseq{1}{M}$ such that 
\begin{align}
\label{eq:th-decoder}
\log \frac{W_{Y|X}(y|x_w)}{Q_Y(y)} > \gamma.
\end{align}
If there is no $w$ satisfying \eqref{eq:th-decoder}, the decoder declares an error. It is known~\cite{Polyanskiy2010} that the conditional probability of error when $W=w$ is upper bounded by $\epsilon_w^{(1)} + \epsilon_w^{(2)}$ where 
\begin{align}
\label{eq:epsilon1}
\epsilon_w^{(1)} &\eqdef  \sum_y W_{Y|X}(y|x_w) \indic{\log \frac{W_{Y|X}(y|x_w)}{Q_Y(y)} \leq \gamma} \editt{=F_{XY|Q_YX=x_w}(\gamma)},\\
\label{eq:epsilon2}
\epsilon_w^{(2)} &\eqdef  \sum_y W_{Y|X}(y|x_w) \indic{\exists w'\neq w:\log \frac{W_{Y|X}(y|x_{w'})}{Q_Y(y)} > \gamma}.
\end{align}
Under random coding, $\epsilon_1^{(1)}, \cdots, \epsilon_w^{(1)}$ are independent, and we can therefore bound their average using well-known concentration inequalities. The following lemma upper bounds the expectation of $\epsilon_w^{(1)}$ and $\epsilon_w^{(2)}$.\todo{Isn't the lemma already in Polyanskiy? We could just cite his paper to save space.}



\begin{lemma} 
\label{lm:ch_coding}
Let $(\mathcal{X}, W_{Y|X}, \mathcal{Y})$ be a \ac{DMC} and $P_X$ and $Q_Y$ be two distributions on $\mathcal{X}$ and $\mathcal{Y}$. \editt{Assume that $P_Y\ll Q_Y$.} If we choose $X_1, \cdots, X_M$ independently according to $P_X$, we have for all $\gamma\in\mathbb{R}$ and $w\in\intseq{1}{M}$
\begin{align}
\E{\epsilon_w^{(1)}}&= F_{XY|Q_Y}(\gamma),\\
\E{\epsilon_w^{(2)}}&\leq \frac{M}{\exp(\gamma)} \E[P_Y]{\frac{P_Y(Y)}{Q_Y(Y)}}.\label{eq:epsilon2-bound}
\end{align}
\end{lemma}

\begin{IEEEproof}
The proof is  similar to the proof of \cite[Lemma 3]{Bloch2016a} or \cite[Theorem 18]{Polyanskiy2010}, but for completeness we provide the short proof. We know that
\begin{align}
\E{\epsilon_w^{(1)}} &= \sum_{x_1, \cdots, x_M}\prod_{k=1}^MP_X(x_k) \sum_y W_{Y|X}(y|x_w)\indic{\log \frac{W_{Y|X}(y|x_w)}{Q_Y(y)} \leq \gamma}\\
&= \sum_{x_w, y} P_X(x_w) W_{Y|X}(y|x_w)\indic{\log \frac{W_{Y|X}(y|x_w)}{Q_Y(y)} \leq \gamma}\displaybreak[0]\\
&= \P[W_{Y|X}P_X]{\log \frac{W_{Y|X}(Y|X)}{Q_Y(Y)}\leq \gamma}\displaybreak[0]\\
&= F_{XY|Q_Y}(\gamma).
\end{align}
Moreover, we have
\begin{align}
\E{\epsilon_w^{(2)}} &\leq \sum_{x_1, \cdots, x_M}\prod_{k=1}^MP_X(x_k) \sum_y W_{Y|X}(y|x_w) \indic{\exists w' \neq w: \log \frac{W_{Y|X}(y|x_{\modified{w'}})}{Q_Y(y)} \geq \gamma}\\
&\leq \sum_{x_1, \cdots, x_M}\prod_{k=1}^MP_X(x_k) \sum_y W_{Y|X}(y|x_w)\sum_{w'\neq w} \indic{\log \frac{W_{Y|X}(y|x_{w'})}{Q_Y(y)} \geq \gamma}\displaybreak[0]\\
&=\sum_{w'\neq w} \sum_{x_1, \cdots, x_M, y} W_{Y|X}(y|x_w)\prod_{k=1}^MP_X(x_k) \indic{\log \frac{W_{Y|X}(y|x_{w'})}{Q_Y(y)} \geq \gamma}\displaybreak[0]\\
&= \sum_{{w'}\neq w} \sum_{x_{w'}, y} P_Y(y) P_X(x_{w'}) \indic{\log \frac{W_{Y|X}(y|x_{w'})}{Q_Y(y)} \geq \gamma}\displaybreak[0]\\
&= \sum_{w'\neq w} \sum_{x_{w'}, y} Q_Y(y) \frac{P_Y(y)}{Q_Y(y)} P_X(x_{w'}) \indic{\log \frac{W_{Y|X}(y|x_{\modified{w'}})}{Q_Y(y)} \geq \gamma}\displaybreak[0]\\
&\leq\sum_{w'\neq w} \sum_{x_{w'}, y}\exp(-\gamma)W_{\modified{Y|X}}(y|x_{w'}) \frac{P_Y(y)}{Q_Y(y)}P_X(x_{w'}) \indic{\log \frac{W_{Y|X}(y|x_{w'})}{\modified{Q}_Y(y)} \geq \gamma}\\
&\leq \frac{M}{\exp(\gamma)} \E[P_Y]{\frac{P_Y(Y)}{Q_Y(Y)}}.
\end{align}
\end{IEEEproof}

\begin{remark}
In a second-order analysis, $\gamma$ is generally chosen such that $\E{\epsilon_w^{(1)}}$ dominates $\E{\epsilon_w^{(2)}}$; we follow this approach as well, which is convenient since we have some flexibility in the bounding of $\E{\epsilon_w^{(1)}}$.
\end{remark}

\begin{remark}
\modified{The extra factor $\E[P_Y]{\frac{P_Y(Y)}{Q_Y(Y)}}$ in \eqref{eq:epsilon2-bound}, which does not appear in  \cite[Theorem 18]{Polyanskiy2010}, is the penalty of using $Q_Y$ instead of the ``true'' probability distribution of $Y$, $P_Y$, in the definition of $F_{XY|Q_Y}$. This sub-optimal bound yields the same first- and second-order asymptotics for covert communications, although it could certainly be improved for a finite-length analysis; this choice , however, simplifies our calculations.}
\end{remark}

Next, we develop one-shot channel resolvability results. In our proofs, we treat the distance between induced output distribution and the desired distribution as a function of independently generated codewords, which allows us to prove a super-exponential concentration inequality in Lemma~\ref{lm:ch_resolve_p} using \modified{McDiarmid}'s Theorem. This may be viewed as an alternative \editt{and more direct} approach to~\cite{Cuff2015} and we recall this concentration inequality below for convenience.

\begin{theorem}[McDiarmid's Theorem]
\label{thm:mcdiarmid}
Let $\textbf{X} \eqdef (X_1, \cdots, X_n)$ be a sequence of independent random variables defined on $\mathcal{X}$. Furthermore, suppose $g:\mathcal{X}^n\to \mathbb{R}$ is a function satisfying
\begin{align}
\sup_{x_1, \cdots, x_n, x_i'}|g(x_1, \cdots, x_i, \cdots, x_n) - g(x_1, \cdots, x_i', \cdots, x_n)| \leq c_i,\quad \forall i \in \intseq{1}{n}.
\end{align} 
Then, for all $\lambda>0$, we have
\begin{align}
\P{g(\textbf{X})-\E{g(\textbf{X})} \geq \lambda} \leq \exp\left(\frac{-2\lambda^2}{\sum c_i^2}\right).
\end{align}
\end{theorem}

\begin{IEEEproof}
 See \cite[Theorem 2.2.2]{Raginsky2014}.
\end{IEEEproof}

\begin{lemma}
\label{lm:ch_resolve_p}
Consider a \ac{DMC} $(\mathcal{X}, W_{Z|X}, \mathcal{Z})$ and a distribution $P_X$ on $\mathcal{X}$. If $\{x_w\}_{w=1}^M \in \mathcal{X}^M$ are \typo{$M\geq 2$} codewords\footnote{\typo{Unlike the main problem in which we have $MK$ codewords, we denote here the total number of codewords by $M$ for simplicity.}}  and \editt{$\widehat{P}_Z(z) {\eqdef \frac{1}{M}\sum_{w=1}^M W_{Z|X}(x|x_w)}$} is the corresponding induced distribution on $\mathcal{Z}$, we define the functions $g_1, g_2:\mathcal{X}^M\to \mathbb{R}$ as
\begin{align}
g_1(x_1, \cdots, x_M) \eqdef \V{\widehat{P}_Z, P_Z},
\end{align}
and
\begin{align}
g_2(x_1, \cdots, x_M) \eqdef \D{\widehat{P}_Z}{P_Z},
\end{align}
with $P_Z(z) = \sum_x P_X(x) W_{Z|X}(z|x)$. Then, for all $x_1, \cdots, x_M, x_i'\in \mathcal{X}$, we have
\begin{align}
|g_1(x_1, \cdots, x_i, \cdots, x_M) - g_1(x_1, \cdots, x_i', \cdots, x_M)| \leq \frac{1}{M},
\end{align}
and
\begin{align}
|g_2(x_1, \cdots, x_i, \cdots, x_M) - g_2(x_1, \cdots, x_i', \cdots, x_M)| \leq \frac{1}{M}\log\left(\frac{M|\mathcal{Z}|}{\mu_{Z}^2}\right).
\end{align}
Moreover, if $X_1, \cdots, X_M$ are \ac{iid} with distribution $P_X$, then
\begin{align}
\label{eq:v-concentration}
\P{\V{\widehat{P}_Z, P_Z}-\E{\V{\widehat{P}_Z,P_Z}}  \geq \lambda} \leq \exp\left(-2M\lambda^2\right),
\end{align}
and
\begin{align}
\label{eq:div-concentration}
\P{\D{\widehat{P}_Z}{P_Z}-\E{\D{\widehat{P}_Z}{P_Z}} \geq \lambda} \leq \exp\left(-\frac{2M\lambda^2}{\log^2\left(\frac{M|\mathcal{Z}|}{\mu_Z^2}\right)}\right).
\end{align}
\end{lemma}

\begin{IEEEproof}
We assume that $\widehat{P}_Z^1$ and $\widehat{P}_Z^2$ are the distributions induced by the codebooks $\mathcal{C}_1\eqdef\{x_1, \cdots, x_i, \cdots, x_M\}$ and  $\mathcal{C}_2\eqdef\{x_1, \cdots, x_i', \cdots, x_M\}$, respectively. First, note that
\begin{align}
g_1(x_1, \cdots, x_i, \cdots, x_M) - g_1(x_1, \cdots, x_i', \cdots, x_M) &= \V{\widehat{P}_Z^1, P_Z} - \V{\widehat{P}_Z^2, P_Z}\\
&\leq \V{\widehat{P}_{Z}^1, \widehat{P}_{Z}^2}\\
&= \frac{1}{2}\sum_{z}\left|\widehat{P}_{Z}^1(z) - \widehat{P}_{Z}^2(z)\right|\\
&=\frac{1}{2}\sum_{z} \left| \frac{1}{M}\sum_{x\in \mathcal{C}_1}W_{Z|X}(z|x)- \frac{1}{M}\sum_{x\in \mathcal{C}_2}W_{Z|X}(z|x) \right|\\
&=\frac{1}{2M}\sum_{z}\left|W_{Z|X}(z|x_i) -W_{Z|X}(z|x_i')\right| \\
&=\frac{1}{M}\V{W_{Z|X}(.|x_i), W_{Z|X}(.|x_i')}\\
&\leq \frac{1}{M}.
\end{align}
Next, for any two distributions $P$ and $Q$, by \cite[Lemma~2.7]{CodingTheoremsDMC2}, we have
\begin{align}
|H(P)-H(Q)| \leq -\V{P, Q}\log \frac{\V{P, Q}}{|\mathcal{X}|}.
\end{align}
Therefore, we have
\begin{align}
\left|H(\widehat{P}_Z^1) - H(\widehat{P}_Z^2)\right| &\leq - \V{\widehat{P}_Z^1, \widehat{P}_Z^2}\log \frac{\V{\widehat{P}_Z^1, \widehat{P}_Z^2}}{|\mathcal{Z}|}\\
&\stackrel{(a)}{\leq} -\frac{1}{M} \log \frac{1}{M|\mathcal{Z}|}\\
&=\frac{1}{M}\log \left(M|\mathcal{Z}|\right),
\end{align}
where $(a)$ follows from the fact that \editt{$x \log (\abs{\mathcal{Z}}/x)$ is increasing for $0 < x < \abs{\mathcal{Z}}/e$}.  Accordingly, we obtain
\begin{align}
\left|\D{\widehat{P}_Z^1}{P_Z}-\D{\widehat{P}_Z^2}{P_Z}\right| 
&= \left|-H(\widehat{P}_Z^1) + \sum_z\widehat{P}_Z^1(z)\log \frac{1}{P_Z(z)} + H(\widehat{P}_Z^2) - \sum_z\widehat{P}_Z^2(z)\log \frac{1}{P_Z(z)}  \right| \\
&\leq \frac{1}{M}\log\left(M|\mathcal{Z}|\right)+\typo{\sum_z\left|\widehat{P}_Z^1(z) - \widehat{P}_Z^2(z)\right|\left|\log \frac{1}{P_Z(z)}\right|}\\
&\leq  \frac{1}{M}\log\left(M|\mathcal{Z}|\right) + \frac{2}{M}\log \frac{1}{\mu_Z}\\
&= \frac{1}{M}\log\left(\frac{M|\mathcal{Z}|}{\mu_Z^2}\right). 
\end{align}
Finally, using Theorem~\ref{thm:mcdiarmid}, we obtain~\eqref{eq:v-concentration} and \eqref{eq:div-concentration}.
\end{IEEEproof}
\editt{The following lemma bounds the expected value of the relative entropy appearing in Lemma~\ref{lm:ch_resolve_p}. The result has appeared in the case of \ac{iid} distributions in~\cite{Hou2014} and we provide here a one-shot result better-suited for covert communications and non-\ac{iid} distributions.}
\begin{lemma}
\label{lm:ch_resolve_e}
Let $(\mathcal{X}, W_{Z|X}, \mathcal{Z})$ be a \ac{DMC} and $P_X$ be a distribution on $\mathcal{X}$. If  $X_1, \cdots, X_M$ are \ac{iid} with distribution $P_X$, and  $\widehat{P}_Z(z) \typo{\eqdef \frac{1}{M} \sum_{w=1}^M W_{Z|X}(z,X_w)}$ is the corresponding induced distribution on $\mathcal{Z}$, we have
\begin{align}
\E{\D{\widehat{P}_Z}{P_Z}} \leq \log \left(\frac{1}{\mu_Z} +1\right)\overline{F}_{XZ}(\gamma)+ \frac{\exp(\gamma)}{M},
\end{align}
for all $\gamma\in\mathbb{R}$ where $P_Z(z) = \sum_x P_X(x) W_{Z|X}(z|x)$.
\end{lemma}

\begin{IEEEproof}
From \cite[Equation 10]{Hou2013}, we know that
\begin{align}
\E{\D{\widehat{P}_Z}{P_Z}} &\leq \E[W_{Z|X}P_X]{\log\left(\frac{W_{Z|X}(Z|X)}{MP_Z(Z)} +1\right)}.
\end{align}
If we define the event $\mathcal{E}\eqdef  \left\{(x, z): \log \frac{W_{Z|X}(z|x))}{P_Z(z)} \leq \gamma\right\}$, then we get
\begin{align}
\E{\log\left(\frac{W_{Z|X}(Z|X)}{MP_Z(Z)} +1\right)} &= \E{\log\left(\frac{W_{Z|X}(Z|X)}{MP_Z(Z)} +1\right)\Bigg|\mathcal{E}}\P{\mathcal{E}} + \E{\log\left(\frac{W_{Z|X}(Z|X)}{MP_Z(Z)} +1\right)\Bigg|\mathcal{E}^c} \P{\mathcal{E}
^c}\\
&\leq \log\left(\frac{\exp(\gamma)}{M} +1\right) +\overline{F}_{XZ}(\gamma) \log \left(\frac{1}{\mu_Z} +1\right)\\
&\leq  \frac{\exp(\gamma)}{M} +\overline{F}_{XZ}(\gamma) \log \left(\frac{1}{\mu_Z} +1\right).
\end{align}
\end{IEEEproof}

Consider now the \ac{DMC} $(\calX, W_{Y|X}, W_{Z|X}, \calY, \calZ)$ and a randomly generated code $\{X_{sw}:s\in \intseq{1}{K},w\in \intseq{1}{M}\}$. The next lemma leverages Lemma~\ref{lm:ch_coding} and Lemma~\ref{lm:ch_resolve_p} to guarantee that, with \editt{positive} probability, the entire code is a channel resolvability code for the channel $W_{Z|X}$ and, \emph{simultaneously}, the subcode corresponding to every $s\in\intseq{1}{K}$ is a channel reliability code for the channel $W_{Y|X}$. Note that a code with these properties is not necessarily a covert code, but we show in the next sections how a careful choice of the random coding distribution leads to covertness.
\editt{
\begin{lemma}
\label{thm:one-shot-covert-cm}
Consider a \ac{DMC} $(\mathcal{X}, W_{Y|X}, W_{Z|X}, \mathcal{Y}, \mathcal{Z})$ and two  probability distributions $P_X$ and $Q_Y$ on $\mathcal{X}$ and $\mathcal{Y}$. We sample codewords $\{X_{sw}:s\in \intseq{1}{K},w\in \intseq{1}{M}\}$ independently according to $P_X$. \editt{For $s\in\intseq{1}{K}$ and $w\in \intseq{1}{M}$, let $\epsilon_{sw}^{(1)}$ and $\epsilon_{sw}^{(2)}$ be defined similar to \eqref{eq:epsilon1} and \eqref{eq:epsilon2}. Moreover, for a fixed subset $\calI \subset \intseq{1}{K}\times \intseq{1}{M}$, we define the induced distribution by the codewords in $\calI$ as $\widehat{P}^{\calI}_{\mathbf{Z}}(z) \eqdef \frac{1}{|\calI|} \sum_{(s, w)\in\calI} W_{Z|X}(z|X_{sw})$.} \typo{The quantities $\epsilon_{sw}^{(1)}$, $\epsilon_{sw}^{(2)}$, and $\D{\widehat{P}_Z^{\calI}}{P_Z}$ will be treated as  random variables with respect to the randomness of the random code}. Finally, for every $\lambda_1$, $\lambda_2$, $\lambda_3$, $\gamma_1$, and $\gamma_2>0$, we define the events
\begin{align}
\label{eq:one-shot-reliability}
{\mathcal{E}_1 \eqdef \left\{\max_{sw} \epsilon^{(1)}_{sw} \leq  \max_{x\in \mathop{\textnormal{supp}} P_X}F_{XY|Q_YX=x}(\gamma_1) \text{ and }  \max_{s} \frac{1}{M} \sum_{w=1}^M \epsilon_{sw}^{(2)} \leq \lambda_2 \frac{M}{ \exp(\gamma_1)}\E[P_Y]{\frac{P_Y(Y)}{Q_Y(Y)}}\right\} }
\end{align}
and
\begin{align}
\label{eq:one-shot-resolvability}
\mathcal{E}_2 \eqdef  \left\{\forall \calI \subset \intseq{1}{K}\times \intseq{1}{M} \text{ with } |\calI| = \lambda_1 MK, \D{\widehat{P}_Z^{\calI}}{P_Z} \leq\log \left(\frac{1}{\mu_Z} +1\right)\overline{F}_{XZ}(\gamma_2)+ \frac{\exp(\gamma_2)}{\lambda_1MK} + \lambda_3\right\}.
\end{align}
Then $\P{\mathcal{E}_1 \cap \mathcal{E}_2} > 0$ if
\begin{align}
\label{eq:mc-criterion}
1 > \exp\left(-M\left(\frac{2\lambda_1\lambda_3^2}{\log^2\left(\frac{\lambda_1MK|\mathcal{Z}|}{\mu_Z^2}\right)} + \Hb{\lambda_1}\right)\right)+\frac{1}{\lambda_2}.
\end{align}
\end{lemma}

\begin{IEEEproof}
We analyze the probability of each event separately. First, note that
\begin{align}
\P{\mathcal{E}_1} &= \P{\max_{sw} \epsilon^{(1)}_{sw} \leq  \max_{x\in \textnormal{supp} P_X}F_{XY|Q_YX=x}(\gamma_1) \text{ and }  \max_{s} \frac{1}{M} \sum_{w=1}^M \epsilon_{sw}^{(2)} \leq \lambda_2 \frac{M}{ \exp(\gamma_1)}\E[P_Y]{\frac{P_Y(Y)}{Q_Y(Y)}}}\displaybreak[0]\\
&\stackrel{(a)}{=} \P{ \max_{s} \frac{1}{M} \sum_{w=1}^M \epsilon_{sw}^{(2)} \leq \lambda_2 \frac{M}{ \exp(\gamma_1)}\E[P_Y]{\frac{P_Y(Y)}{Q_Y(Y)}}}\displaybreak[0]\\
& \stackrel{(b)}{=}\left(1-\P{\frac{1}{M}\sum_{w=1}^M\epsilon_{1w}^{(2)} >\lambda_2 \frac{M}{ \exp(\gamma_1)}\E[P_Y]{\frac{P_Y(Y)}{Q_Y(Y)}}}\right)^K\displaybreak[0]\\
&\stackrel{(c)}{\geq} \left(1-\frac{1}{\lambda_2}\right)^K,
\end{align}
where $(a)$ follows since by \eqref{eq:epsilon1}, $\max_{sw} \epsilon^{(1)}_{sw} \leq  \max_{x\in \textnormal{supp} P_X}F_{XY|Q_YX=x}(\gamma_1) $ holds almost surely, $(b)$ follows since all sub-codebooks are generated independently, and $(c)$ follows from Markov inequality and Lemma~\ref{lm:ch_coding}.  Furthermore, we have
\begin{align}
  \P{\mathcal{E}_2} &= \mathbb{P}\left(\forall \calI \subset \intseq{1}{K}\times \intseq{1}{M} \text{ with } |\calI| = \lambda_1 MK, \D{\widehat{P}_Z^{\calI}}{P_Z} \leq\log \left(\frac{1}{\mu_Z} +1\right)\overline{F}_{XZ}(\gamma_2)\right.\nonumber\\
  &\left.\phantom{========================================}+ \frac{\exp(\gamma_2)}{\lambda_1MK} + \lambda_3\right)  \\ 
&\geq 1-\sum_{\calI\subset\intseq{1}{K}\times \intseq{1}{M}: |\calI| = \lambda_1 MK} \P{\D{\widehat{P}_Z^{\calI}}{P_Z} \geq\log \left(\frac{1}{\mu_Z} +1\right)\overline{F}_{XZ}(\gamma_2)+ \frac{\exp(\gamma_2)}{\lambda_1MK} + \lambda_3}\\
&\stackrel{(a)}{\geq} 1-\sum_{\calI\subset\intseq{1}{K}\times \intseq{1}{M}: |\calI| = \lambda_1 MK} \P{\D{\widehat{P}_Z^{\calI}}{P_Z} \geq\E{\D{\widehat{P}_Z^{\calI}}{P_Z}} + \lambda_3} \\ 
&\stackrel{(b)}{\geq} 1-  \binom{MK}{\lambda_1MK} \exp\left(-\frac{2MK\lambda_1\lambda_3^2}{\log^2\left(\frac{\lambda_1MK|\mathcal{Z}|}{\mu_Z^2}\right)}\right)\\
&\stackrel{(c)}{\geq} 1 - \exp\left(MK\Hb{\lambda_1}\right)\exp\left(-\frac{2MK\lambda_1\lambda_3^2}{\log^2\left(\frac{\lambda_1MK|\mathcal{Z}|}{\mu_Z^2}\right)}\right),
\end{align}
where $(a)$ follows from Lemma~\ref{lm:ch_resolve_e}, $(b)$ follows from Lemma~\ref{lm:ch_resolve_p}, and $(c)$ follows from $\binom{MK}{\lambda_1 MK}\leq \exp\left(MK\Hb{\lambda_1}\right)$. Combining these inequalities, we get
\begin{align}
\label{eq:mc-events}
\P{\mathcal{E}_1} + \P{\mathcal{E}_2} \geq\left(1 -\frac{1}{\lambda_2}\right)^K + 1 -\exp\left(MK\Hb{\lambda_1}\right) \exp\left(-\frac{2MK\lambda_1\lambda_3^2}{\log^2\left(\frac{\lambda_1MK|\mathcal{Z}|}{\mu_Z^2}\right)}\right).
\end{align}
Thus, we obtain
\begin{align}
\P{\mathcal{E}_1\cap \mathcal{E}_2} &= 1- \P{\mathcal{E}_1^c\cup \mathcal{E}_2^c}\\
&\geq 1-\P{\mathcal{E}_1^c}-\P[\mathbb{C}]{\mathcal{E}_2^c}\\
&=\P{\mathcal{E}_1}+\P{\mathcal{E}_2}-1\\
&\stackrel{(a)}{\geq}\left(1-\frac{1}{\lambda_2}\right)^K  - \exp\left(MK\Hb{\lambda_1}\right)\exp\left(-\frac{2MK\lambda_3^2}{\log^2\left(\frac{MK|\mathcal{Z}|}{\mu_Z^2}\right)}\right), \label{eq:p-one-covert}
\end{align}
where $(a)$ follows from  \eqref{eq:mc-events}. Finally, \eqref{eq:mc-criterion} ensures that the RHS of \eqref{eq:p-one-covert} is positive.
\end{IEEEproof}
}

\subsection{Asymptotic results for covert communications}
\label{sec:asympt-results-cover}

We specialize the one-shot result of Lemma~\ref{thm:one-shot-covert-cm} to show the existence of covert codes for large blocklength $n$.
\editt{
\begin{theorem}
\label{thm:general_achv}
Consider a sequence of distributions $\{P_\textbf{X}^n\}_{n\geq 1}$ where $P_\textbf{X}^n$ is defined on $\calX^n$ and two sequences of natural numbers $\{M_n\}_{n\geq 1}$ and $\{K_n\}_{n\geq 1}$. For a quasi-metric $d$ and $n$ large enough, if we have
\begin{align}
\label{eq:positive_prob}
 1 - \exp\left(-M_n\left(\frac{2(1-n^{-11})}{n^{8}\log^2\left(\frac{M_nK_n|\mathcal{Z}|^n}{(1-n^{-11})\mu_Z^{2n}}\right)} - \Hb{{n^{-11}}}\right)\right)-\frac{1}{n} > 0,
\end{align}
\begin{align}
\label{eq:F_Y}
 \max_{\mathbf{x}\in \textnormal{supp} P_{\mathbf{X}}^n}F_{\mathbf{X}\mathbf{Y}|P_0^\pn\mathbf{X}=\mathbf{x}}(n^{13}\log M_n) +  \frac{1}{n}\E[P_Y]{\frac{P_{\mathbf{Y}}(\mathbf{Y})}{P_0^\pn(\mathbf{Y})}}\leq \epsilon ,
\end{align}
\begin{align}
\label{eq:F_Z}
\overline{F}_{\textbf{X}\mathbf{Z}}\left(\log \frac{M_n K_n}{n^4}\right) \leq n^{-6},
\end{align}
and
\begin{align}
\label{eq:dist_P_Z}
d(P_{\mathbf{Z}}, Q_0^\pn) \leq \delta - n^{-\frac{1}{2}},
\end{align}
then $M_d^*(n, \epsilon, \delta) \geq M_n$.
\end{theorem}
Theorem~\ref{thm:general_achv} is useful in that it identifies sufficient conditions for the existence of codes. In particular, subsequent achievability proofs are reduced to choosing an appropriate sequence of random coding distributions $\{P_\textbf{X}^n\}_{n\geq 1}$ and verifying inequalities~(\ref{eq:positive_prob})-(\ref{eq:dist_P_Z}). We \editt{will use a} \ac{PPM} distribution, which we introduce and analyze precisely in Section~\ref{sec:distributions}.


\begin{IEEEproof}
For a fixed $n$, we choose codewords $\textbf{X}_{sw}$ for $s\in\intseq{1}{K_n}$ and $w\in \intseq{1}{M_n}$ independently according to $P_\textbf{X}^n$. In Lemma~\ref{thm:one-shot-covert-cm}, if we set 
\begin{align}
\lambda_1 = 1-n^{-11},\quad \lambda_2 = n, \quad, \lambda_3 = n^{-4},\quad, \gamma_1 = \log \left(n^{13} M_n\right), \quad \text{and } \gamma_2 = \log \frac{M_n K_n}{n^4},
\end{align} 
then  \eqref{eq:positive_prob} implies \eqref{eq:mc-criterion} for large $n$. Therefore, we have with positive probability  for $n$ large enough,
 \begin{align}
\max_{s,w} \epsilon^{(1)}_{sw} &\leq  \max_{\mathbf{x}\in \textnormal{supp} P_{\mathbf{X}}^n}F_{\mathbf{X}\mathbf{Y}|P_0^\pn\mathbf{X}=\mathbf{x}}(\gamma_1),
 \label{eq:asym_reliable}\\
  \max_{s} \frac{1}{M_n} \sum_{w=1}^{M_n} \epsilon_{sw}^{(2)} &\leq \lambda_2 \frac{M_n}{ \exp(\gamma_1)}\E[P_Y]{\frac{P_{\mathbf{Y}}(\mathbf{Y})}{P_0^\pn(\mathbf{Y})}},
 \end{align}
where \eqref{eq:asym_reliable} follows from~\eqref{eq:epsilon1} with $Q_Y = P_0^\pn$. Also, for all $\calI\subset\intseq{1}{K_n}\times \intseq{1}{M_n}$ with $|\calI| = \lambda_1 M_nK_n$ and with $\mu_{\mathbf{Z}} \eqdef \min_\textbf{z} P_{\mathbf{Z}}(\textbf{z})$, we have for large enough $n$
 \begin{align}
 \D{\widehat{P}_{\mathbf{Z}}^\calI}{P_{\mathbf{Z}}} &\leq \log\left(1 + \frac{1}{\mu_\mathbf{Z}}\right)\overline{F}_{\textbf{X}\mathbf{Z}}(\gamma_2)+ \frac{\exp(\gamma_2)}{\lambda_1M_nK_n} + \lambda_3\\
 \label{eq:asym_resolve}
 &\stackrel{(a)}{\leq} n \log\left(1 + \frac{1}{\mu_Z}\right) \frac{1}{n^6} + \frac{1}{n^4} + \frac{1}{n^4}\leq \frac{3}{n^4},
 \end{align}
 where $(a)$ follows from \eqref{eq:F_Z} and the choice of $\gamma_2$, $\lambda_3$. We next expurgate the $(1-\lambda_1)M_n$ codewords with largest $\epsilon_{sw}^{(2)}$ for all $s$ from the sub-codebook  $\{\mathbf{X}_{sw}: w\in\intseq{1}{M_n}\}$ to form a new codebook $\calI$ with $|\calI| = \lambda_1 M_nK_n$. For the new code, by Markov inequality, we have
\begin{align}
 \max_{s, w} \epsilon_{sw}^{(2)} \leq \frac{1}{1-\lambda_1}\lambda_2 \frac{M_n}{ \exp(\gamma_1)}\E[P_Y]{\frac{P_{\mathbf{Y}}(\mathbf{Y})}{P_0^\pn(\mathbf{Y})}}.
\end{align} 
Hence, the maximum probability of error \editt{for the expurgated code is} bounded by
\begin{align}
\max_{s, w} (\epsilon_{sw}^{(1)}+\epsilon_{sw}^{(2)})
& \leq   \max_{\mathbf{x}\in \textnormal{supp} P_{\mathbf{X}}^n}F_{\mathbf{X}\mathbf{Y}|P_0^\pn\mathbf{X}=\mathbf{x}}(\gamma_1) +  \frac{1}{1-\lambda_1}\lambda_2 \frac{M_n}{ \exp(\gamma_1)}\E[P_Y]{\frac{P_{\mathbf{Y}}(\mathbf{Y})}{P_0^\pn(\mathbf{Y})}}\\
&=  \max_{\mathbf{x}\in \textnormal{supp} P_{\mathbf{X}}^n}F_{\mathbf{X}\mathbf{Y}|P_0^\pn\mathbf{X}=\mathbf{x}}(\gamma_1) +  \frac{1}{n}\E[P_Y]{\frac{P_{\mathbf{Y}}(\mathbf{Y})}{P_0^\pn(\mathbf{Y})}}\leq \epsilon.
\end{align}
  Finally, to prove the covertness of the code, note that for $n$ large enough
\begin{align}
d(\widehat{P}_{\mathbf{Z}}^\calI, Q_0^\pn) &\stackrel{(a)}{\leq} d(P_\mathbf{Z}^n, Q_0^\pn) + \D{\widehat{P}^\calI_{\mathbf{Z}}}{P_\mathbf{Z}^n} + \sqrt{ \D{\widehat{P}_{\mathbf{Z}}^\calI}{P_\mathbf{Z}^n}}\max\left(1,\log \frac{1}{\min_{\mathbf{z}:Q_0^\pn(\mathbf{z})>0} Q_0^\pn(\mathbf{z})}\right)\\
&\stackrel{(b)}{\leq} \delta - \frac{1}{\sqrt{n}}+ \D{\widehat{P}^\calI_{\mathbf{Z}}}{P_\mathbf{Z}^n} + \sqrt{ \D{\widehat{P}^\calI_{\mathbf{Z}}}{P_\mathbf{Z}^n}}\max\left(1,\log \frac{1}{\min_{\mathbf{z}:Q_0^\pn(\mathbf{z})>0} Q_0^\pn(\mathbf{z})}\right)\\
&\stackrel{(c)}{\leq}  \delta - \frac{1}{\sqrt{n}}+\frac{3}{n^4} + \sqrt{ \frac{3}{n^4}}\max\left(1,\log \frac{1}{\min_{\mathbf{z}:Q_0^\pn(\mathbf{z})>0} Q_0^\pn(\mathbf{z})}\right)\\
&\leq  \delta - \frac{1}{\sqrt{n}}+\frac{3}{n^4} + \sqrt{ \frac{3}{n^4}}\max\left(1, n\log \frac{1}{\mu_Z}\right)\\
 &\leq \delta,
\end{align}
where $(a)$ follows because $d$ is a quasi-metric, $(b)$ follows from \eqref{eq:dist_P_Z}, and $(c)$ follows from \eqref{eq:asym_resolve}.
\end{IEEEproof}}

\subsection{PPM distribution}
\label{sec:distributions}

\editt{We now }study \editt{the} specific ``\ac{PPM} distribution," which was used in~\cite{Bloch2016b} for covert communication; we subsequently use this distribution to generate random codes and combine with Theorem~\ref{thm:general_achv}. \editt{The choice of \ac{PPM} is crucial to evaluate~\eqref{eq:F_Y}.}


\begin{definition}[$(n, \ell)$-\ac{PPM} covert distribution]
\label{def:ppm_distribution}
Given $\mathcal{X} = \{0, 1\}$ and $n\geq \ell \geq 1$, we define \editt{the} distribution $\ppmX{n}{\ell}$ on $\mathcal{X}^n$ \editt{as follows}. If $n = m\ell + r$ for $0\leq r < \ell$, we partition the set $\intseq{1}{n}$ into $\ell$ sets of size $m$ and one set of size $r$. For simplicity, we consider the following partition: $\mathcal{B}_i = \{(i-1)m+1, \cdots, im\}$ for $i\in\intseq{1}{\ell}$ and $\mathcal{B}_{\ell+1} = \{\ell m+1, \cdots, n\}$. Then, $P_{\mathbf{X},\textnormal{PPM}}^{n, \ell}$ is the uniform distribution on the sequences with exactly one ``1" in all $\mathcal{B}_i$ for $i\in\intseq{1}{\ell}$ and no ``1" in $\mathcal{B}_{\ell+1}$, i.e.,
\begin{align}
\left\{\mathbf{x}\in\mathcal{X}^n: \forall i \in \intseq{1}{\ell}: \sum_{j\in\mathcal{B}_i}x_j = 1~ \mathrm{ and } \sum_{j\in \mathcal{B}_{\ell+1}} x_j= 0\right\}.
\end{align}
We denote the output distribution of $P_{\mathbf{X}, \textnormal{PPM}}^{n,\ell}$ for \ac{DMC}s $(\mathcal{X}, W_{Y|X}, \mathcal{Y})$ and $(\mathcal{X}, W_{Z|X}, \mathcal{Z})$ by $P_{\mathbf{Y}, \textnormal{PPM}}^{n, \ell}$ and $P_{\mathbf{Z}, \textnormal{PPM}}^{n, \ell}$, respectively.
\end{definition}

The \ac{PPM} covert distribution $\ppmX{n}{\ell}$ may be viewed as a way to generate structured constant composition codewords; specifically, every codeword generated according to $\ppmX{n}{\ell}$ contains exactly $\ell$ 1-symbols, each located in a window of size $m$, the quotient of the Euclidean division of $n$ by $\ell$. The structure of the \ac{PPM} covert distribution plays a pivotal role in our analysis.

We devote the remainder of this section to the development of properties of PPM distribution, which are geared towards the use of Theorem~\ref{thm:general_achv}. We make repeated use of the Berry-Esseen Theorem, which we recall here for convenience.
\begin{theorem}[Berry-Esseen Theorem]
\label{Berry-Esseen}
Let $X_1, \cdots, X_n$ be independent random variables s.t. for $k\in\intseq{1}{n}$ we have $ \E{X_k} = \mu_k$, $\sigma_k^2=\Var{X_k}$, and $t_k=\E{|X_k-\mu_k|^3}$.
If we define $\sigma^2 = \sum_{\modified{k}=1}^\modified{n} \sigma_k^2$ and $T=\sum_{k=1}^n t_\modified{k}$, then we have
\begin{align}
 \left|\P{\sum_{k=1}^n(X_k-\mu_k) \geq \lambda \sigma}-Q(\lambda)\right| \leq \frac{6T}{\sigma^3}.
\end{align}
\end{theorem}	

We \editt{now upper bound} the \ac{lhs} of~\eqref{eq:F_Y} for the \ac{PPM} distribution in Lemma~\ref{lm:F_XY_bound} and Lemma~\ref{lm:expected_ration_ppm}.
\editt{
\begin{lemma}
\label{lm:F_XY_bound}
Let $(\mathcal{X}, W_{Y|X}, \mathcal{Y})$ be a binary-input \ac{DMC}. 
If $(\mathbf{X}, \mathbf{Y})$ is distributed according to $\ppmX{n}{\ell}W_{Y|X}^\pn$, we have
\begin{align}
 \max_{\mathbf{x}\in \textnormal{supp}\ppmX{n}{\ell}}F_{\mathbf{X}\mathbf{Y}|P_0^\pn\mathbf{X}=\mathbf{x}}(\gamma) \leq Q\left(\frac{\ell D_P - \gamma}{\sqrt{\ell V_P}}\right) + \frac{6T_P}{\sqrt{\ell}V_P^{\frac{3}{2}}},
\end{align}
with
\begin{align}
T_P \eqdef \E[P_1]{\left|\log\frac{P_1(Y)}{P_0(Y)} - \D{P_1}{P_0} \right|^3}.
\end{align}
\end{lemma}

\begin{IEEEproof}
For a sequence $\mathbf{x}_0$  with $\ppmX{n}{\ell}(\textbf{x}_0) > 0$, we can assume without loss of generality that the support of $\mathbf{x}_0$ is $\intseq{1}{\ell}$ since the channel is memoryless and by relabeling. Hence, we have
\begin{align}
F_{\mathbf{X}\mathbf{Y}|P_0^\pn\mathbf{X}=\mathbf{x}_0}(\gamma)
&=\P[W_{Y|X=\mathbf{x}_0}^\pn]{\log \frac{W_{Y|X}^\pn(\mathbf{Y}|\textbf{x}_0)}{P_0^\pn(\mathbf{Y})}\leq \gamma | \mathbf{X}=\mathbf{x}_0}\\
&= \P[P_1^{\proddist \ell}]{\sum_{i=1}^\ell\log \frac{P_1(Y_i)}{P_0(Y_i)}\leq \gamma}\\
&\stackrel{(a)}{\leq}  Q\left(\frac{\ell D_P - \gamma}{\sqrt{\ell V_P}}\right) + \frac{6T_P}{\sqrt{\ell}V_P^{\frac{3}{2}}},
\end{align}
where $(a)$ follows from Theorem~\ref{Berry-Esseen}.
%
\end{IEEEproof}
}
\begin{lemma}
\label{lm:expected_ration_ppm}
For a binary-input \ac{DMC} $(\mathcal{X}, W_{Y|X}, \mathcal{Y})$, we have
\begin{align}
\label{eq:ratio_ppm}
\E[\ppmY{n}{\ell}]{\frac{\ppmY{n}{\ell}(\mathbf{Y})}{P_0^\pn(\mathbf{Y})}} &= \left(1 + \frac{\chi_2(P_1\|P_0)}{\lfloor n /\ell \rfloor}\right)^{\ell}\\
&\leq \exp\left(\frac{\ell(\ell+1)}{n}\chi_2(P_1\|P_0)\right).
\end{align}
\end{lemma}

\begin{IEEEproof}
To prove \eqref{eq:ratio_ppm}, we define $m\eqdef \lfloor n /\ell \rfloor$. We first consider \typo{$\ppmX{m}{1}$} for which we have
\begin{align}
\E[\typo{\ppmY{m}{1}}]{\frac{\typo{\ppmY{m}{1}}(\mathbf{Y})}{P_0^{\proddist m}(\mathbf{Y})}} 
&= \sum_\textbf{y} \typo{\ppmY{m}{1}}(\textbf{y})\frac{\typo{\ppmY{m}{1}}(\textbf{y})}{P_0^{\proddist m}(\textbf{y})}\\
&= \sum_\textbf{y} P_0^{\proddist m}(\textbf{y})\left(\frac{\typo{\ppmY{m}{1}}(\textbf{y})}{P_0^{\proddist m}(\textbf{y})}\right)^2\\
&= \sum_\textbf{y} P_0^{\proddist m}(\textbf{y}) \left(\frac{1}{m}\sum_{i=1}^m \frac{P_1(y_i)}{P_0(y_i)}\right)^2\\
&= \frac{1}{m^2}\left(m(m-1)+m(\chi_2(P_1\|P_0) +1)\right)\\
&= 1 + \frac{\chi_2(P_1\|P_0)}{m}.
\end{align}
Since $\ppmY{n}{\ell}$ is the product of the distributions over different blocks, i.e., $\ppmY{n}{\ell} = \left(\ppmY{m}{1}\right) ^ {\proddist \ell} \proddist \typo{P}_0^{\proddist n \text{ mod } \ell}$, upon defining $\mathbf{Y}_i = (Y_{(i-1)m+1}, \cdots, Y_{im})$ for $i\in\intseq{1}{\ell}$, we obtain
\begin{align}
\E[\ppmY{n}{\ell}]{\frac{\ppmY{n}{\ell}(\mathbf{Y})}{P_0^\pn(\mathbf{Y})}} &= \E[\left(\ppmY{m}{1}\right) ^ {\proddist \ell}]{\prod_{i=1}^\ell \frac{\ppmY{m}{1}(\mathbf{Y}_i)}{P_0^{\proddist m}(\mathbf{Y}_i)}}\\
&= \prod_{i=1}^\ell\E[\ppmY{m}{1}]{\frac{\ppmY{m}{1}(\mathbf{Y}_i)}{P_0^{\proddist m}(\mathbf{Y}_i)}}\\
&= \left(1 + \frac{\chi_2(P_1\|P_0)}{m}\right)^{\ell}\\
&\leq \exp\left(\frac{\ell(\ell+1)}{n}\chi_2(P_1\|P_0)\right).
\end{align}
\end{IEEEproof}
We now upper-bound the \ac{lhs} of~\eqref{eq:F_Z} in Lemma~\ref{lm:F_XZ_bound}.
\begin{lemma}
\label{lm:F_XZ_bound}
Given a binary-input \ac{DMC} $(\mathcal{X}, W_{Z|X}, \mathcal{Z})$,
  if $(\mathbf{X}, \mathbf{Z})$ is distributed according to $\ppmX{n}{\ell}W_{Z|X}^\pn$, then, for $\gamma \geq \ell D_Q$, we have
\begin{align}
\overline{F}_{\mathbf{X}\mathbf{Z}}(\gamma) \leq \exp\left(-\frac{\left(\gamma-\ell D_Q\right)^2}{2\ell \log^2\mu_Z}\right).
\end{align}
\end{lemma}

\begin{IEEEproof}
When $(\mathbf{X}, \mathbf{Z})$ is distributed according to $\ppmX{n}{\ell}W_{Z|X}^\pn$, for the random vector $\mathbf{Z}\in \mathcal{Z}^n$, the first $\ell$ blocks of length $m\eqdef \lfloor n/\ell \rfloor$ are denoted by $\mathbf{Z}_1, \cdots, \mathbf{Z}_{\ell}$ with $\mathbf{Z}_i = (Z_{(i-1)m+1}, \cdots, Z_{im})$. Moreover,  $P_{\mathbf{Z}_i}$ denotes the distribution of block $\mathbf{Z}_i$. Therefore, we have
\begin{align}
\overline{F}_{\mathbf{X}\mathbf{Z}}(\gamma)&= \P[\ppmX{n}{\ell}W_{Z|X}^\pn]{\log \frac{W_{Z|X}^\pn(\mathbf{Z}|\mathbf{X})}{\ppmZ{n}{\ell}(\mathbf{Z})}\geq \gamma}\\
&= \P[\ppmX{n}{\ell}W_{Z|X}^\pn]{\sum_{i=1}^\ell \log \frac{W_{Z|X}^{\proddist m}(\mathbf{Z}_i|\mathbf{X}_i)}{P_{\mathbf{Z}_i}(\mathbf{Z}_i)}\geq \gamma}\\
&\stackrel{(a)}{\leq} \exp\left(-\frac{2\left(\gamma-\ell I(\ppmX{m}{1}, W_{Z|X}^{\proddist m})\right)^2}{\ell B^2}\right),
\end{align}
where $(a)$ follows from Hoeffding's Inequality with
\begin{align}
B\eqdef \sup_{\textbf{x}_i, \textbf{z}_i}\left| \log \frac{W_{Z|X}^{\proddist m}(\textbf{z}_i|\textbf{x}_i)}{P_{\mathbf{Z}_i}(\textbf{z}_i)}\right| > 0
\end{align}
and $I(\ppmX{m}{1}, W_{Z|X}^{\proddist m})$ is defined as in \editt{\eqref{eq:I_def}}.
By \cite[Lemma~1.5]{Bloch2016b}, we know that $I(\ppmX{m}{1}, W_{Z|X}^{\proddist m}) \leq D_Q$ and \editt{$B \leq \log m$ for $m$ large enough}.\footnote{The bound for $B$ corrects a small oversight with no incidence on the result in~\cite{Bloch2016b}.} Hence, we obtain
\begin{align}
\exp\left(-\frac{2\left(\gamma-\ell I(\ppmX{m}{1}, W_{Z|X}^{\proddist m})\right)^2}{\ell B^2}\right) \leq  \exp\left(-\frac{\left(\gamma-\ell D_Q\right)^2}{\ell\log^2m}\right).
\end{align}
\end{IEEEproof}
\modified{
  Finally, we upper-bound the \ac{lhs} of~\eqref{eq:dist_P_Z} using Lemma~\ref{lm:div_ppm}.}
\begin{lemma}
\label{lm:div_ppm}
Let  $(\calX, W_{Z|X}, \calZ)$ be a \ac{DMC}, and $n$ and $\ell$ be positive integers with $m \eqdef \left \lfloor n/\ell \right \rfloor$ large enough and $\ell = \Theta(m)$. Then, we have
\begin{align}
\D{\ppmZ{n}{\ell}}{Q_0^\pn} &\leq \frac{\ell^2}{2n}\chi_2(Q_1\|Q_0) + O\pr{\frac{1}{\sqrt{n}}}\label{eq:d_ppm}.
\end{align}
\modified{Furthermore, we have
\begin{align}
\V{\ppmZ{n}{\ell},Q_0^\pn} &\leq 1- 2Q\pr{\frac{\ell}{2}\sqrt{\frac{\chi_2(Q_1\|Q_0)}{n}}} + \frac{2}{\sqrt{\ell}} + O\pr{\frac{1}{\sqrt{n}}}\label{eq:v_ppm},\\
\beta_\alpha\left(Q_0^\pn, \ppmZ{n}{\ell}\right) &\geq Q\pr{\ell \sqrt{\frac{\chi_2(Q_1\|Q_0)}{n}} - Q^{-1}\pr{\alpha + \frac{1}{\sqrt{\ell}}}} - \frac{1}{\sqrt{\ell}} + O\pr{\frac{1}{\sqrt{n}}}.\label{eq:beta_ppm}
\end{align}}
\end{lemma}

\begin{IEEEproof}
By \cite[Lemma 1]{Bloch2016b}, we have for \editt{$n$ large enough} and some positive constant $C_1$, $C_2$, $C_3$, \editt{and $C_4$,}
\begin{align}
 \D{\ppmZ{n}{\ell}}{Q_0^\pn} &\leq \frac{\ell}{2m} \chi_2(Q_1\|Q_0) + \frac{\ell C_1}{m^2} + \frac{\ell C_2}{m^3}\\
 &\leq \frac{\ell}{2m} \chi_2(Q_1\|Q_0) + \frac{\ell C_3}{m^2}\\
 &\leq \frac{\ell}{2(n/\ell - 1)} \chi_2(Q_1\|Q_0) + \frac{\ell C_3}{(n/\ell - 1)^2}\\
 &\leq \frac{\ell^2}{2n}\chi_2(Q_1\|Q_0) + \frac{\ell^3C_4}{n^2}.
\end{align}

\modified{To prove \eqref{eq:v_ppm}, note that for any two distributions $P$ and $Q$, one can write $\V{P, Q} = \P[P]{P(X) \geq Q(X)} - \P[Q]{P(X) \geq Q(X)}$. Therefore, if $\mathbf{Z} = (Z_1, \cdots, Z_n)$ and $\widetilde{\mathbf{Z}} = (\widetilde{Z}_1, \cdots, \widetilde{Z}_n)$ are distributed according to $Q_0^\pn$ and $\ppmZ{n}{\ell}$, respectively, then
\begin{align}
\V{\ppmZ{n}{\ell}, Q_0^\pn} 
&= \P{\ppmZ{n}{\ell}(\widetilde{\mathbf{Z}}) \geq Q_0^\pn(\widetilde{\mathbf{Z}})} - \P{\ppmZ{n}{\ell}({\mathbf{Z}}) \geq Q_0^\pn({\mathbf{Z}})}\\
&= \P{\log \frac{\ppmZ{n}{\ell}(\widetilde{\mathbf{Z}})}{Q_0^\pn(\widetilde{\mathbf{Z}})} \geq 0} - \P{\log \frac{\ppmZ{n}{\ell}({\mathbf{Z}}) }{ Q_0^\pn({\mathbf{Z}})} \geq 0}.
\end{align}
However, note that $\ppmZ{n}{\proddist \ell} = (\ppmZ{m}{1})^\ell \otimes Q_0^{\proddist r}$, and for any $\mathbf{z} = (z_1, \cdots, z_n)$ with $\mathbf{z}_i = (z_{(i-1)m + 1},\cdots, z_{im})$, we have
\begin{align}
\log\pr{ \frac{\ppmZ{n}{\ell}({\mathbf{z}})}{Q_0^\pn({\mathbf{z}})} }
&= \sum_{i=1}^\ell \log\pr{ \frac{\ppmZ{m}{1}(\mathbf{z}_i)}{Q_0^{\proddist m}(\mathbf{z}_i)}}\\
&= \sum_{i=1}^\ell \log\pr{ \frac{\sum_{j=1}^m \frac{1}{m} Q_1(z_{(i-1)m + j}) \prod_{k\neq j} Q_0(z_{(i-1)m + k})}{Q_0^{\proddist m}(\mathbf{z}_i)}}\\
&= \sum_{i=1}^\ell \log \pr{\frac{1}{m}\sum_{j=1}^m\frac{  Q_1(z_{(i-1)m + j})}{ Q_0(z_{(i-1)m + j})}}\\
&= \sum_{i=1}^\ell \log \pr{1 + \frac{1}{m}\sum_{j=1}^m\frac{  Q_1(z_{(i-1)m + j})-Q_0(z_{(i-1)m + j})}{ Q_0(z_{(i-1)m + j})}}\\
&= \sum_{i=1}^\ell \log \pr{1 + \frac{1}{m}\sum_{j=1}^mA(z_{(i-1)m+j})},
\end{align}
for $A(z) \eqdef \frac{Q_1(z) - Q_0(z)}{Q_0(z)}$. Thus, if we define 
\begin{align}
B_i &\eqdef \frac{1}{m} \sum_{j=1}^m A(Z_{(i-1)m + j})\\
C_i &\eqdef \log(1+B_i)\\
\widetilde{B}_i &\eqdef \frac{1}{m} \sum_{j=1}^m A(\widetilde{Z}_{(i-1)m + j})\\
\widetilde{C}_i &\eqdef \log(1+\widetilde{B}_i).
\end{align}
Then, we have
\begin{align}
\P{\log \frac{\ppmZ{n}{\ell}(\widetilde{\mathbf{Z}})}{Q_0^\pn(\widetilde{\mathbf{Z}})} \geq 0} - \P{\log \frac{\ppmZ{n}{\ell}({\mathbf{Z}}) }{ Q_0^\pn({\mathbf{Z}})} \geq 0}
&= \P{\sum_{i=1}^\ell \widetilde{C}_i \geq 0} - \P{\sum_{i=1}^\ell C_i \geq 0}.
\end{align}
Furthermore, we consider two \ac{iid} sequences $\underline{B}_1,  \cdots, \underline{B}_\ell$ and $\widetilde{\underline{B}}_1,  \cdots, \widetilde{\underline{B}}_\ell$  with distribution 
\begin{align}
\P{\underline{B}_i = b} &\eqdef \P{B_i = b|B_i \neq -1}\\
\P{\widetilde{\underline{B}}_i = b} &\eqdef \P{\widetilde{B}_i = b|\widetilde{B}_i \neq -1}.
\end{align}
Note that with $\tau\eqdef \P{A(Z_1) = -1} < 1$, we have $\P{B_i = -1} = \tau^m$. By defining $\underline{C}_i \eqdef \log(1 + \underline{B}_i)$ and $\underline{\widetilde{C}}_i \eqdef \log(1+\widetilde{\underline{B}}_i)$, we will deal with random variables that take finite values with probability one. We \editt{now} replace $\widetilde{C}_i$ by $\underline{\widetilde{C}}_i$, and since $\underline{\widetilde{C}}_1, \cdots  \underline{\widetilde{C}}_{\ell}$ are \ac{iid}, we apply Theorem~\ref{Berry-Esseen} to obtain
\begin{align}
 \P{\sum_{i=1}^\ell \widetilde{C}_i \geq 0} 
  &=  \P{\sum_{i=1}^\ell \widetilde{C}_i \geq 0\big| \forall i, \widetilde{B}_i \neq -1} \P{\forall i, \widetilde{B}_i \neq -1}\nonumber\\
  &\phantom{=============} + \P{\sum_{i=1}^\ell \widetilde{C}_i \geq 0\big| \exists i: \widetilde{B}_i = -1} \P{\exists i: \widetilde{B}_i = -1} \displaybreak[0]\\
 &\leq \P{\sum_{i=1}^\ell \widetilde{C}_i \geq 0\big| \forall i, \widetilde{B}_i \neq -1} + \ell \tau^m\displaybreak[0]\\
 &= \P{\sum_{i=1}^\ell \widetilde{\underline{C}}_i \geq 0} + O\pr{\frac{1}{\sqrt{n}}}\\
 &\leq Q\pr{\frac{- \sum_{i=1}^\ell \E{\widetilde{\underline{C}}_i}}{\sqrt{\ell \sum_{i=1}^\ell \Var{\widetilde{\underline{C}}_i}}}}  + \frac{\sum_{i=1}^\ell \E{\left|\widetilde{\underline{C}}_i - \E{\widetilde{\underline{C}}_i}\right|^3}}{\pr{\sum_{i=1}^\ell \Var{\widetilde{\underline{C}}_i}}^{\frac{3}{2}}}\displaybreak[0]\\
 &\stackrel{(a)}{\leq} Q\pr{-\frac{\ell \pr{ \frac{\chi_2(Q_1\|Q_0)}{2m} + O\pr{\frac{1}{m^2}}}}{\sqrt{\ell\pr{\frac{\chi_2(Q_1\|Q_0)}{m} + O\pr{\frac{1}{m^2}}} }}} + \frac{\ell\pr{\frac{\chi_2(Q_1\|Q_0)^{\frac{3}{2}}}{m^{\frac{3}{2}}} + O\pr{\frac{1}{m^{\frac{9}{4}}}}}}{\pr{\ell\pr{\frac{\chi_2(Q_1\|Q_0)}{m} + O\pr{\frac{1}{m^2}}}}^{\frac{3}{2}}}\displaybreak[0]\\
 &= Q\pr{-\frac{1}{2} \sqrt{\frac{\ell \chi_2(Q_1\|Q_0)}{m} }} + \frac{1}{\sqrt{\ell}} + O\pr{\frac{1}{\sqrt{n}}},
\end{align}
where $(a)$ follows from Lemma~\ref{lm:A-moments} in Appendix~\ref{sec:technical-lemma}. Analogously, using Lemma~\ref{lm:A-moments} we obtain
\begin{align}
\P{\sum_{i=1}^\ell {C}_i \geq 0}  \geq Q\pr{\frac{1}{2} \sqrt{\frac{\ell \chi_2(Q_1\|Q_0)}{m} }} - \frac{1}{\sqrt{\ell}} + O\pr{\frac{1}{\sqrt{n}}},\label{eq:q0-log-ratio}
\end{align}
and therefore, we have \eqref{eq:v_ppm}.

To lower-bound $\beta_{\alpha}( Q_0^\pn, \ppmZ{n}{\ell})$, we use the Neyman-Pearson lemma, which states that if  $\alpha \leq \P{\log \frac{\ppmZ{n}{\ell}({\mathbf{Z}})}{Q_0^\pn({\mathbf{Z}})} \geq \gamma}$ for some $\gamma$, then we have $\beta_{\alpha}( Q_0^\pn, \ppmZ{n}{\ell}) \geq \P{\log \frac{\ppmZ{n}{\ell}(\widetilde{\mathbf{Z}})}{Q_0^\pn(\widetilde{\mathbf{Z}})} \leq \gamma}$ where $\mathbf{Z}$ and $\widetilde{\mathbf{Z}}$ are defined as before. Using Theorem~\ref{Berry-Esseen} again, similar to \eqref{eq:q0-log-ratio}, we obtain
\begin{align}
\P{\log \frac{\ppmZ{n}{\ell}({\mathbf{Z}})}{Q_0^\pn({\mathbf{Z}})} \geq \gamma} \geq Q\pr{\frac{\gamma + \frac{\ell \chi_2(Q_1\|Q_0)}{2m}}{\sqrt{\frac{\ell \chi_2(Q_1\|Q_0)}{m}}}} - \frac{1}{\sqrt{\ell}} + O\pr{\frac{1}{\sqrt{n}}}.
\end{align}
Accordingly, to ensure that $\P{\log \frac{\ppmZ{n}{\ell}({\mathbf{Z}})}{Q_0^\pn({\mathbf{Z}})} \leq \gamma} \geq \alpha$, we choose $\gamma = \sqrt{\frac{\ell \chi_2(Q_1\|Q_0)}{m}} Q^{-1}\pr{\alpha + \frac{1}{\sqrt{\ell}}} -\frac{\ell\chi_2(Q_1\|Q_0)}{2m} + O\pr{\frac{1}{\sqrt{n}}}$. Thus, we have
\begin{align}
\beta_{\alpha}( Q_0^\pn, \ppmZ{n}{\ell}) 
&\geq \P{\log \frac{\ppmZ{n}{\ell}(\widetilde{\mathbf{Z}})}{Q_0^\pn(\widetilde{\mathbf{Z}})} \leq \sqrt{\frac{\ell \chi_2(Q_1\|Q_0)}{m}} Q^{-1}\pr{\alpha + \frac{1}{\sqrt{\ell}}} -\frac{\ell\chi_2(Q_1\|Q_0)}{2m} + O\pr{\frac{1}{\sqrt{n}}}}\\
&\geq Q\pr{\frac{-\sqrt{\frac{\ell \chi_2(Q_1\|Q_0)}{m}} Q^{-1}\pr{\alpha + \frac{1}{\sqrt{\ell}}} +\frac{\ell\chi_2(Q_1\|Q_0)}{2m} + \frac{\ell \chi_2(Q_1\|Q_0)}{2m}}{\sqrt{\frac{\ell \chi_2(Q_1\|Q_0)}{m}}}} - \frac{1}{\sqrt{\ell}} + O\pr{\frac{1}{\sqrt{n}}}\\
&= Q\pr{\sqrt{\frac{\ell \chi_2(Q_1\|Q_0)}{m}} - Q^{-1}\pr{\alpha + \frac{1}{\sqrt{\ell}}}} - \frac{1}{\sqrt{\ell}} + O\pr{\frac{1}{\sqrt{n}}}.
\end{align}}
\end{IEEEproof}

\subsection{Converse}
\label{sec:generic-converse}
We now develop a generic converse for a quasi-metric $d$. The following result states that the number of covert and reliable bits that one can transmit may be characterized by establishing an upper bound on the weight of codewords. Such upper bounds are metric-specific, and we develop them in Section~\ref{sec:covert-comm-with}. 
\begin{theorem}
\label{thm:converse-general}
\editt{ Let $d$ be a quasi-metric and $(\calX, W_{Y|X}, W_{Z|X}, \calY, \calZ)$ be a \ac{DMC}. If there exists  \modified{ a function $g:\mathbb{R} \to \mathbb{R}$} such that for every $(M, K, n, \epsilon, \delta)_d$ code $\mathcal{C}$, there exists a subset of the codewords, $\mathcal{D}$, with
 \begin{align}
 \label{eq:wt-condition-metric}
 \frac{\max_{\textbf{x} \in\mathcal{D}} \wt{\textbf{x}}}{\sqrt{n}} \leq g(\delta) + \frac{C}{\sqrt{n}},
 \end{align}
 and $|\calD| \geq \frac{MK}{n^h}$ where $C, h>0$ only depend on the channel, then
 \begin{align}
 \log M^*_d(n, \epsilon, \delta) \leq g(\delta)D_P n^{\frac{1}{2}} - \sqrt{g(\delta)V_P} Q^{-1}(\epsilon)n^{\frac{1}{4}} + O(\log n).
 \end{align}
In addition, if for every $(M_n, K_n, n, \epsilon, \delta)_d$ code in a sequence $\{\calC_n\}$ of increasing length $n$ such that $\lim_{n\rightarrow\infty}\epsilon_n=0$, there exists a non-empty subset of codewords $\mathcal{D}_n\subset\calC_n$ with 
\begin{align}
\label{eq:wt-condition-metric-first-order}
\frac{\max_{\textbf{x} \in\mathcal{D}_n} \wt{\textbf{x}}}{\sqrt{n}} \leq g(\delta) + o(1),\quad \text{ and } \log\frac{M_nK_n}{\card{\calD_n}}=o\left(\sqrt{n}\right).
\end{align}
Then,
\begin{align}
  \lim_{\epsilon\rightarrow 0}\lim_{n\rightarrow\infty} \frac{\log M^*_d(n,\epsilon,\delta)}{\sqrt{n}}=g(\delta).
\end{align}}
\end{theorem}

\begin{IEEEproof}
\editt{
Let $\mathcal{C}$ be an $(M, K, n, \epsilon, \delta)_d$ code. For $s\in \intseq{1}{K}$, we denote the sub-codebook of all codewords characterized by the key value $s$ by $\mathcal{C}^s$. By our assumptions, we can choose $\mathcal{D} \subset \mathcal{C}$ of size at least $MK/n^h$ satisfying \eqref{eq:wt-condition-metric}. By the pigeonhole principle, there should be at least one sub-codebook $\mathcal{C}^s$ such that $|\mathcal{D}\cap \mathcal{C}^s| \geq M / n^h$. Let us define
\begin{align}
\mathcal{D}^s_i \eqdef \{\mathbf{x}\in \mathcal{D}\cap \mathcal{C}^s: \wt{\textbf{x}} = i\}.
\end{align} 
Since $\mathcal{D}^s_i$ is included in $\mathcal{C}^s$, it is a reliability code for the channel $W_{Y|X}$ with maximum probability of error less than or equal to $\epsilon$. Moreover, the type of the codewords in $\mathcal{D}^s_i$ is fixed, and therefore for any $\textbf{x}\in\mathcal{D}_i^s$,   \cite[Theorem 28]{Polyanskiy2010} yields that
\begin{align}
\log |\mathcal{D}^s_i| &\leq -\log \beta_{1-\epsilon}\left(W_{\mathbf{Y}|\mathbf{X} = \textbf{x}}, P_0^\pn\right). 
\end{align}
Furthermore, similar to \cite{Yang2016}, by \cite[Equation 102]{Polyanskiy2010}, we obtain for any $\gamma > 0$
\begin{align}
-\typo{\log} \beta_{1-\epsilon}\left(W_{\mathbf{Y}|\mathbf{X} = \textbf{x}}, P_0^\pn\right) \leq \gamma - \log \left(1-\epsilon - \P[W_{\mathbf{Y}|\mathbf{X} = \textbf{x}}]{\log \frac{W_{\mathbf{Y}|\mathbf{X}}(\mathbf{Y}|\textbf{x})}{P_0^\pn(\mathbf{Y})} \geq \gamma}\right).
\end{align}
However, we know that $W_{\mathbf{Y}|\mathbf{X}}(\mathbf{Y}|\textbf{x}) = \prod_{k: x_k = 1} P_1(Y_k) \prod_{j:x_j = 0} P_0(Y_j)$ which yields that if $\tilde{Y}_1, \cdots, \tilde{Y}_i$ are \ac{iid} according to $P_1$, then
\begin{align}
 \P[W_{\mathbf{Y}|\mathbf{X} = \textbf{x}}]{\log \frac{W_{\mathbf{Y}|\mathbf{X}}(\mathbf{Y}|\textbf{x})}{P_0^\pn(\mathbf{Y})} \geq \gamma} = \P[P_1^{\proddist i}]{\sum_{k=1}^i \log \frac{P_1(\tilde{Y}_k)}{P_0(\tilde{Y}_k)} \geq \gamma}.
\end{align}
By Theorem~\ref{Berry-Esseen}, we have
\begin{align}
\P{\sum_{k=1}^i \log \frac{P_1(\tilde{Y}_k)}{P_0(\tilde{Y}_k)} \geq \gamma} \leq Q\left(\frac{\gamma - i D_P}{\sqrt{iV_P}}\right) + \frac{B}{\sqrt{i}},
\end{align}
where $B$ just depends on the channel. Setting $\gamma = iD_P + \sqrt{iV_P} Q^{-1}(1-\epsilon-\frac{2B}{\sqrt{i}})$ and combining all above equations, we have
\begin{align}
\log |\mathcal{D}_i^s| \leq i D_P + \sqrt{iV_P}Q^{-1}(1-\epsilon) -\log \left(\frac{B}{\sqrt{i}}\right).
\end{align}
Hence, we obtain
\begin{align}
\log\frac{M}{n^h} &\leq \log |\mathcal{D} \cap \mathcal{C}^s| \\
&= \log \left(\sum_{i=0}^{\sqrt{n} g(\delta) + C} |\mathcal{D}_i^s| \right)\\
&\leq \log \left(\sum_{i=0}^{\sqrt{n} g(\delta) + C}\exp( i D_P + \sqrt{iV_P}Q^{-1}(1-\epsilon) -\log \left(\frac{B}{\sqrt{i}})\right)\right)\\
&\leq \log\left( (\sqrt{n}g(\delta)+C) \exp(g(\delta)D_Pn^{\frac{1}{2}}-\sqrt{V_Pg(\delta) }Q^{-1}(\epsilon)n^{\frac{1}{4}}+ O(1)) \right).
\end{align}
Thus, we get
\begin{align}
\log M \leq g(\delta)D_P n^{\frac{1}{2}}-\sqrt{g(\delta)V_P}Q^{-1}(\epsilon)n^{\frac{1}{4}}+(h+1)\log n - \log(B) + \log   (\sqrt{n}g(\delta)+C) + O(1).
\end{align}
}
\editt{
  Note that we need not worry about exotic channels for $\epsilon\in(\frac{1}{2},1)$ since the constants $D_P$ and $V_P$ are known and fixed.

  Consider now a sequence of  $(M_n,K_n,n,\epsilon_n,\delta)$ code for which subsets $\calD_n\subset \calC_n$ exist and~\eqref{eq:wt-condition-metric-first-order} holds. Recall that $\epsilon_n$ is defined as per~\eqref{eq:prob_error}. By the pigeon hole principle, for every $n$, there exists a sub-codebook $\calC_n(s)$ consisting of all codewords indexed by the same key $s\in\intseq{1}{K_n}$, such that $\card{\calC_n(s)\cap\calD_n}\geq \frac{\card{\calD_n}}{K_n}$, and the maximum probability of error corresponding to $\calC_n(s)\cap\calD_n$ is at most $\epsilon_n$.
Following the exact same converse steps as in~\cite[Section VI]{Bloch2016a}, we obtain for $n$ large enough that
\begin{align}
  \log \card{\calC_n(s)\cap\calD_n} \leq \frac{1}{1-\epsilon_n}\left(\sqrt{n}\left(g(\delta)+o(1)\right) + \Hb{\epsilon_n}\right). 
\end{align}
Consequently,
\begin{align}
  \log M_n &\leq \log \card{\calC_n(s)\cap\calD_n} + \log\frac{M_nK_n}{\card{\calD_n}}\\
           &\leq \frac{1}{1-\epsilon_n}\left(\sqrt{n}\left(g(\delta)+o(1)\right) + \Hb{\epsilon_n}\right)+\log\frac{M_nK_n}{\card{\calD_n}}
\end{align}
and
\begin{align}
  \lim_{n\rightarrow\infty} \frac{\log M_n}{\sqrt{n}} = g(\delta). 
\end{align}
}
\end{IEEEproof}

\section{Covert Communication with specific covertness metrics}
\label{sec:covert-comm-with}

We now leverage the general results established in Section~\ref{sec:covert-comm-results} and specialize them to study three covertness metrics: relative entropy (Subsection~\ref{sec:covert-comm-with-kl}), variational distance (Subsection~\ref{sec:secon-order-covert-tv}), and probability of missed detection (Subsection~\ref{sec:covert-comm-missed-detection}). As alluded to earlier, all that needs to be done is: (i) establish that the metric under consideration is a quasi metric, as defined at the beginning of Section~\ref{sec:covert-comm-results}; (ii) verify that the conditions of Theorem~\ref{thm:general_achv} and Theorem~\ref{thm:converse-general} are satisfied.

\subsection{Covertness in relative entropy}
\label{sec:covert-comm-with-kl}
In this subsection, we prove Theorem~\ref{th:main_result_d}.
\begin{lemma}
Relative entropy is a quasi-metric, i.e., 
\begin{align}
\forall P, Q, R:\quad \D{R}{ Q} \leq \D{P}{ Q} + \D{R}{P}+\sqrt{\D{R}{P}}\max\left(1,\log \frac{1}{\min_{z:Q(z)>0} Q(z)}\right).
\end{align}
\end{lemma}

\begin{IEEEproof}
Note that
\begin{align}
\D{R}{ Q} &= \sum_z R(z) \log \frac{R(z)}{Q(z)}\displaybreak[0]\\
&=  \sum_z R(z) \log \frac{R(z)}{P(z)}+ \sum_z R(z) \log \frac{P(z)}{Q(z)}\displaybreak[0]\\
&= \D{R}{P} + \sum_z R(z) \log \frac{P(z)}{Q(z)}\displaybreak[0]\\
&= \D{R}{P} + \D{P}{Q} + \sum_z (R(z) - P(z))\log \frac{P(z)}{Q(z)}\displaybreak[0]\\
&\leq \D{R}{P} + \D{P}{Q} + \log \frac{1}{\min_{z:Q(z)>0} Q(z)} \V{R, P}\displaybreak[0]\\
&\leq \D{R}{P} + \D{P}{Q} + \sqrt{\D{R}{ P}}\log \frac{1}{\min_{z:Q(z)>0} Q(z)}\displaybreak[0]\\
&\leq \D{P}{ Q} + \D{R}{P}+\sqrt{\D{R}{P}}\max\left(1,\log \frac{1}{\min_{z:Q(z)>0} Q(z)}\right).
\end{align}
\end{IEEEproof}

\begin{IEEEproof}[\typo{Achievability of Theorem~\ref{th:main_result_d}}]
Fix $\epsilon \in ]0, 1[$ and $\delta > 0$, and define $\omega \eqdef \sqrt{\frac{2\delta}{\chi_2(Q_1\|Q_0)}}$ and $\ell_n \eqdef  \lfloor \omega \sqrt{n} -t \rfloor$ where the value of $t$ will be determined later.  To use Theorem~\ref{thm:general_achv}, we choose $P_\mathbf{X}^n \eqdef \ppmX{n}{\ell_n}$, and we set
\begin{align}
\log M_ n &\eqdef  \ell_n D_P - \sqrt{\ell_n V_P}Q^{-1}\left(\epsilon - \frac{1+6T_P/V_P^{\frac{3}{2}}}{\sqrt{\ell_n} }\right) - \editt{13}\log n\\
&= \omega D_P n ^{\frac{1}{2}} - \sqrt{\omega V_P}Q^{-1}\left(\epsilon\right)n^{\frac{1}{4}} - \editt{13}\log n + O(1),
\end{align}
and
\begin{align}
\log M_n + \log K_n &\eqdef \max(\log M_n, (1+\rho) \ell_n D_Q)\\
&= \max(\log M_n, (1+\rho) \omega D_Q n^{\frac{1}{2}} + O(1)).
\end{align}
 By Lemma~\ref{lm:div_ppm}, we know that for \editt{$n$ large enough and some $C>0$}
\begin{align}
\D{\ppmZ{n}{\ell_n}}{Q_0^\pn} &\leq \frac{\ell_n^2}{2n}\chi_2(Q_1\|Q_0) + \frac{\ell_n^3C}{n^2}\\
&\leq \frac{(\omega\sqrt{n} - t)^2}{2n}\chi_2(Q_1\|Q_0) + \frac{(\omega\sqrt{n} - t)^3C}{n^2}\\
&\leq \frac{1}{2}\omega^2\chi_2(Q_1\|Q_0) - \frac{t\omega \chi_2(Q_1\|Q_0)}{\sqrt{n}} + \frac{t^2\chi_2(Q_1\|Q_0)}{2n} + \frac{\omega^3C}{\sqrt{n}}\\
&\stackrel{(a)}{\leq} \frac{1}{2}\omega^2\chi_2(Q_1\|Q_0) - \frac{1}{\sqrt{n}}, 
\end{align}
where $(a)$ is true for \editt{$t > \frac{1 + \omega^3 C}{\omega \chi_2(Q_1\|Q_0)}$} and $n$ large enough. Furthermore, by Lemma~\ref{lm:expected_ration_ppm}, we have
\begin{align}
\E[\ppmY{n}{\ell_n}]{\frac{\ppmY{n}{\ell_n}(\mathbf{Y})}{P_0^\pn(\mathbf{Y})}}&\leq \exp\left(\frac{\ell_n(\ell_n+1)}{n}\chi_2(P_1\|P_0)\right) = O(1).
\end{align}
Hence, for large enough $n$, we have
\editt{\begin{multline}
F_{\mathbf{X}\mathbf{Y}|P_0^\pn}\left(\log \left(n^{13}M_n\right)\right) +\frac{\E[\ppmY{n}{\ell_n}]{\frac{\ppmY{n}{\ell_n}(\mathbf{Y})}{P_0^\pn(\mathbf{Y})}}}{n}\\
\begin{split}
 &= F_{\mathbf{X}\mathbf{Y}|P_0^\pn}\left( \ell_n D_P - \sqrt{\ell_n V_P}Q^{-1}\left(\epsilon - \frac{1+6T_P/V_P^{\frac{3}{2}}}{\sqrt{\ell_n} }\right) \right)  + O\left(\frac{1}{n}\right)\\
&\stackrel{(a)}{\leq}Q\left(\frac{\ell_n D_P -  \ell_n D_P + \sqrt{\ell_n V_P}Q^{-1}\left(\epsilon - \frac{1+6T_P/V_P^{\frac{3}{2}}}{\sqrt{\ell_n} }\right) }{\sqrt{\ell_nV_P}}\right) + \frac{6T_P}{\sqrt{\ell_n V_P^3}} + O\left(\frac{1}{n}\right)\\
&\leq \epsilon,
\end{split}\label{eq:bound_f_xy_div}
\end{multline}}
where $(a)$ follows from Lemma~\ref{lm:F_XY_bound}. By Lemma~\ref{lm:F_XZ_bound}\editt{, for any $\rho > 0$, }we have
\begin{align}
\overline{F}_{\mathbf{X}\mathbf{Z}}\left(\log \frac{M_nK_n}{n^4}\right) &\leq \exp\left(-\frac{\left(\log \frac{M_nK_n}{n^4}-\ell_n D_Q\right)^2}{\ell_n\log^2(n/\ell_n)}\right)\displaybreak[0]\\
&\leq\exp\left(-\frac{\left( (1+\rho) \ell _n D_Q - 4\log n-\ell_n D_Q\right)^2}{\ell_n\log^2(n/\ell_n)}\right)\displaybreak[0]\\
&\leq\exp\left(-\frac{\ell_n\left( \rho D_Q - 4\log n / \ell_n \right)^2}{\log^2(n/\ell_n)}\right)\\
&\stackrel{(a)}{\leq} \frac{1}{n^6},
\end{align}
where $(a)$ is true for large enough $n$. Thus, all the conditions in Theorem~\ref{thm:general_achv} hold, and we obtain
\begin{align}
\log M^*_D(n, \epsilon, \delta) 
  \geq \omega D_P n ^{\frac{1}{2}} - \sqrt{\omega V_P}Q^{-1}\left(\epsilon\right)n^{\frac{1}{4}} - \editt{13}\log n + O(1).
\end{align}
\end{IEEEproof}

\begin{IEEEproof}[\typo{Converse of Theorem~\ref{th:main_result_d}}]
\modified{\editt{We only focus on the converse for second-order asymptotics since the first-order result follows from~\cite{Bloch2016a,Wang2016b}}. For an $(M, K, n, \epsilon, \delta)_D$ code, we \editt{briefly outline} for completeness the argument of \cite{Bloch2016a} to obtain an upper-bound on the average weight of the codewords. Note that, by \cite[Equation (96)]{Bloch2016a},
\begin{align}
\delta \geq \D{\widehat{P}_{\mathbf{Z}}}{Q_0^n}
\stackrel{(a)}{\geq} n\D{\overline{P}_Z}{Q_0},
\end{align}
where $\overline{P}_Z(z) \eqdef \frac{1}{n} \widehat{P}_{Z_i}(z)$. We know that \editt{$\overline{P}_Z = \mu Q_{1} + (1-\mu)Q_0$} for $\mu \eqdef \frac{1}{MK} \sum_{w\in\intseq{1}{M}, s\in \intseq{1}{K}} \frac{\wt{\mathbf{x}_{s, w}}}{n}$. Therefore, by \cite[Equation (11)]{Bloch2016a}, we have
\begin{align}
\D{\overline{P}_Z}{Q_0} \geq \frac{\mu^2}{2}\chi_2(Q_1\|Q_0) - O(\mu^3).
\end{align}
Combining these two bounds, we obtain
\begin{align}
\frac{1}{MK}\sum_{s=1}^K\sum_{w=1}^M \wt{\textbf{x}_{sw}}\leq (B+\sqrt{n})\sqrt{\frac{2\delta}{\chi_2(Q_1\|Q_0)}},
\end{align}
for some $B>0$ depending on the channel. \editt{Thus, we can choose a subset $\calD$ of codewords of size $MK/n$ such that
\begin{align}
\frac{\max_{\textbf{x}\in\mathcal{D}} \wt{\textbf{x}}}{\sqrt{n}} \leq \frac{n}{n-1}\left(\frac{B}{\sqrt{n}}+1\right)\sqrt{\frac{2\delta}{\chi_2(Q_1\|Q_0)}}.
\end{align}}
Applying Theorem~\ref{thm:converse-general} completes the converse proof.}
\end{IEEEproof}

\subsection{Covertness with variational distance}
\label{sec:secon-order-covert-tv}
In this subsection, we prove Theorem~\ref{th:main_result_tv}.
\begin{lemma}
Total variation is a quasi-metric for distributions, i.e., for all distributions $P$, $Q$, and $R$ defined over \modified{$\calZ$}, we have
\begin{align}
\V{R, Q} \leq \V{P, Q} + \D{R}{Q} + \sqrt{\D{R}{P}}\max\left(1,\log \frac{1}{\min_{\modified{z}:Q(z)>0} Q(\modified{z})}\right).
\end{align}
\end{lemma}

\begin{IEEEproof}
Total variation is a metric, and therefore we have
\begin{align}
\V{R, Q} & \leq \V{P, Q} + \V{R, P}\\
&\stackrel{(a)}{\leq} \V{P, Q} + \sqrt{\D{R}{P}}\\
&\leq \V{P, Q} + \D{R}{Q} + \sqrt{\D{R}{P}}\max\left(1,\log \frac{1}{\min_{\modified{z}:Q(z)>0} Q(\modified{z})}\right),
\end{align}
where $(a)$ follows from Pinsker's Inequality.
\end{IEEEproof}

\begin{IEEEproof}[\typo{Achievability of Theorem~\ref{th:main_result_tv}}]
\modified{Fix $\epsilon, \delta \in ]0, 1[$ and $n$. Similar to the analysis for relative entropy, we use Theorem~\ref{thm:general_achv} with $P_\mathbf{X}^n \eqdef \ppmX{n}{\ell_n}$ where $\ell_n$ is chosen to ensure $\V{\ppmZ{n}{\ell_n}, Q_0^\pn}\leq \delta - \frac{1}{\sqrt{n}}$. Specifically, upon setting $\Gamma \eqdef Q^{-1}\pr{\frac{1-\delta}{2}}$, we wish to choose $\ell_n$ such that
\begin{align}
\label{eq:ell_bound1}
\frac{\ell_n}{2} \sqrt{\frac{\chi_2(Q_1\|Q_0)}{n}} \geq \Gamma -\frac{e^{\frac{\Gamma^2}{2}} \sqrt{2\pi}}{\sqrt{\ell_n}} + O\pr{\frac{1}{\sqrt{n}}}.
\end{align}
The term $\Gamma$ is what we would expect by inspection of~(\ref{eq:v_ppm}) to obtain a variational distance with first order approximately $\delta$. As we shall see next, the penalty term $\frac{e^{\frac{\Gamma^2}{2}} \sqrt{2\pi}}{\sqrt{\ell_n}}$  is the result of a Taylor series of the $Q$ function around $\Gamma$. Formally, we choose $\sqrt{\ell_n}$ as a solution of the cubic equation
\begin{align}
x^3 - 2\pr{\Gamma + O\pr{\frac{1}{\sqrt{n}}}}\sqrt{\frac{n}{\chi_2(Q_1\|Q_0)}} x + 2\sqrt{2\pi} e^{\frac{\Gamma^2}{2}} \sqrt{\frac{n}{\chi_2(Q_1\|Q_0)}} = 0,
\end{align}
which will be sufficient to ensure~\eqref{eq:ell_bound1}. For a cubic equation of the form $x^3 - px + q = 0$, the roots are known in closed algebraic form and one of them is $2\sqrt{\frac{-p}{3}}\cos\pr{\frac{1}{3}\arccos\pr{\frac{3q}{2p} \sqrt{\frac{-3}{p}}}}$. Thus,
\begin{align}
\sqrt{\ell_n} 
&= 2\sqrt{\frac{2\pr{\Gamma + O\pr{\frac{1}{\sqrt{n}}}}\sqrt{n}}{3\sqrt{\chi_2(Q_1\|Q_0)}}}\cos\pr{\frac{1}{3}\arccos\pr{-\frac{3\sqrt{2\pi} e^{\frac{\Gamma^2}{2}}}{2\pr{\Gamma + O\pr{\frac{1}{\sqrt{n}}}}} \sqrt{\frac{3\sqrt{\chi_2(Q_1\|Q_0)}}{2\pr{\Gamma+O\pr{\frac{1}{\sqrt{n}}}}\sqrt{n}}}}}\displaybreak[0]\\
&\stackrel{(a)}{=}2\sqrt{\frac{2\pr{\Gamma + O\pr{\frac{1}{\sqrt{n}}}}\sqrt{n}}{3\sqrt{\chi_2(Q_1\|Q_0)}}}\cos\pr{\frac{1}{3}\pr{\frac{\pi}{2} + \frac{3\sqrt{2\pi} e^{\frac{\Gamma^2}{2}}}{2\Gamma } \sqrt{\frac{3\sqrt{\chi_2(Q_1\|Q_0)}}{2\Gamma\sqrt{n}}} + O\pr{\frac{1}{\sqrt{n}}}}}\displaybreak[0]\\
&\stackrel{(b)}{=}  2\sqrt{\frac{2\pr{\Gamma + O\pr{\frac{1}{\sqrt{n}}}}\sqrt{n}}{3\sqrt{\chi_2(Q_1\|Q_0)}}} \pr{\frac{\sqrt{3}}{2} -\frac{1}{2}\frac{\sqrt{2\pi} e^{\frac{\Gamma^2}{2}}}{2\Gamma } \sqrt{\frac{3\sqrt{\chi_2(Q_1\|Q_0)}}{2\Gamma\sqrt{n}}} + O\pr{\frac{1}{\sqrt{n}}}}\\
&= \sqrt{\frac{2\Gamma\sqrt{n}}{\sqrt{\chi_2(Q_1\|Q_0)}}} - \sqrt{\frac{\pi}{2}}\frac{ e^{\frac{\Gamma^2}{2}}}{\Gamma} + O\pr{\frac{1}{n^{\frac{1}{4}}}},\label{eq:sq_ell_v}
\end{align}
where $(a)$ and $(b)$ follow since, for $x$ close to zero,  $\arccos(x) = \frac{\pi}{2} - x + O(x^2)$ and $\cos\pr{\frac{\pi}{6} + x} = \frac{\sqrt{3}}{2} - \frac{1}{2}x + O(x^2)$, respectively. Consequently, upon choosing 
\begin{align}
\ell_n = 2 \Gamma \sqrt{\frac{n}{\chi_2(Q_1\|Q_0)}} - \frac{2\sqrt{\pi} e^{\frac{\Gamma^2}{2}} n^{\frac{1}{4}}}{\sqrt{\Gamma} \chi_2(Q_1\|Q_0)^{\frac{1}{4}}}  + O\pr{1},
\end{align}
we satisfy \eqref{eq:ell_bound1}. Combining with \eqref{eq:v_ppm}, we have
\begin{align}
\V{\ppmZ{n}{\ell_n}, Q_0^\pn} 
&\leq 1-2Q\pr{\frac{\ell_n}{2} \sqrt{\frac{\chi_2(Q_1\|Q_0)}{n}}} + \frac{2}{\sqrt{\ell_n}} + O\pr{\frac{1}{\sqrt{n}}}\\
&\leq 1-2Q\pr{\Gamma -\frac{e^{\frac{\Gamma^2}{2}} \sqrt{2\pi}}{\sqrt{\ell_n}} + O\pr{\frac{1}{\sqrt{n}}}} + \frac{2}{\sqrt{\ell_n}}+ O\pr{\frac{1}{\sqrt{n}}}\\
&\stackrel{(a)}{\leq} 1 - 2Q(\Gamma) - \frac{2}{\sqrt{\ell_n}} + \frac{2}{\sqrt{\ell_n}} + O\pr{\frac{1}{\sqrt{n}}}\\
&= \delta + O\pr{\frac{1}{\sqrt{n}}},  \label{eq:v_bound_ppm}
\end{align}
where $(a)$ follows from $Q(x - y) \geq Q(x) + \frac{e^{-\frac{x^2}{2}}y}{\sqrt{2\pi}}$ for $x \geq y \geq 0$. Finally, we can control the term $O\pr{\frac{1}{\sqrt{n}}}$ in \eqref{eq:ell_bound1} to guarantee that \eqref{eq:v_bound_ppm} is less than $\delta - \frac{1}{\sqrt{n}}$. 

Finally, let
\begin{align}
\log M_ n &\eqdef  \ell_n D_P - \sqrt{\ell_n V_P}Q^{-1}\left(\epsilon - \frac{1+6T_P/V_P^{\frac{3}{2}}}{\sqrt{\ell_n} }\right) - \editt{13}\log n
\end{align}
and for some $\rho > 0$,
\begin{align}
\log M_n + \log K_n &\eqdef \max(\log M_n, (1+\rho) \ell_n D_Q).
\end{align}
Similar to~\eqref{eq:bound_f_xy_div}, we can show that, for large enough $n$,
\editt{
\begin{align}
F_{\mathbf{X}\mathbf{Y}|P_0^\pn}\left(\log \left(n^{\editt{13}}M_n\right)\right) +\frac{\E[\ppmY{n}{\ell_n}]{\frac{\ppmY{n}{\ell_n}(\mathbf{Y})}{P_0^\pn(\mathbf{Y})}}}{n} \leq \epsilon,
\end{align}}
and 
\begin{align}
\overline{F}_{\mathbf{X}\mathbf{Z}}\left(\log \frac{M_nK_n}{n^4}\right)  \leq \frac{1}{n^6}.
\end{align}
Thus, all the conditions in Theorem~\ref{thm:general_achv} hold, and we obtain
\begin{align}
  \log M^*_V(n, \epsilon, \delta) 
  &\geq \ell_n D_P - \sqrt{\ell_n V_P}Q^{-1}\left(\epsilon\right) - \editt{13}\log n + O(1)\\
&=  \frac{2\Gamma D_Pn^{\frac{1}{2}}}{\sqrt{\chi_2(Q_1\|Q_0)}} - \pr{\sqrt{ \frac{2\Gamma  V_P}{\sqrt{\chi_2(Q_1\|Q_0)}}}Q^{-1}\left(\epsilon\right) +\frac{2\sqrt{\pi}e^{\frac{\Gamma^2}{2}}D_P}{\sqrt{\Gamma} \chi_2(Q_1\|Q_0)^{\frac{1}{4}}} }n^{\frac{1}{4}}- \editt{13}\log n + O(1).
\end{align}}
\end{IEEEproof}

To develop the converse for variational distance, we start by relating the variational distance $\V{\widehat{P}_\mathbf{Z}, Q_0^{\pn}} $ to the minimum weight of the codewords.
\begin{lemma}
\label{lm:converse-wmin}
Consider a binary-input \ac{DMC} $(\calX, W_{Z|X}, \calZ)$ and $M$ codewords $\textbf{x}_1, \cdots, \textbf{x}_M\in\mathcal{X}^n$ with induced distribution $\widehat{P}_\mathbf{Z}$ on $\mathcal{Z}^n$. If $w_{\min} \eqdef \min_{m\in \intseq{1}{M}}\wt{\textbf{x}_m}$, then we have
\begin{align}
\V{\widehat{P}_\mathbf{Z}, Q_0^{\pn}} \geq 1 - 2Q\left(\frac{w_{\min}\sqrt{\chi_2(Q_1\|Q_0)}}{2\sqrt{n} }\right)- \frac{B}{\sqrt{n}} - \frac{w_{\min}^2B}{n^{\frac{3}{2}}},
\end{align}
where $B$ is a constant that only depends on the channel.
\end{lemma}

\begin{IEEEproof}
To lower bound  $\V{\widehat{P}_\mathbf{Z}, Q_0^{\pn}}$, we introduce \editt{a} hypothesis testing problem with two hypotheses $H_0$ and $H_1$ corresponding to distributions  $Q_0^{\pn}$ and $\widehat{P}_\mathbf{Z}$, respectively. We know that for any test with probability of false alarm and missed detection $\alpha$ and $\beta$, respectively, we have
\begin{align}
\label{eq:tv_ht}
\V{\widehat{P}_{\mathbf{Z}}, Q_0^{\pn}} \geq 1 - \alpha - \beta.
\end{align}
Hence, to link the variational distance to the weight of codewords, it suffices to introduce a test for which $\alpha$ and $\beta$ conveniently relate to the weight of codewords. We consider here the sub-optimal test
\begin{align}
T(\textbf{z}) \eqdef \indic{\sum_{i=1}^n A(z_i) > \tau},
\end{align}
where $A(z) \eqdef \frac{Q_1(z) - Q_0(z)}{Q_0(z)}$ and $\tau$ is an arbitrary constant that is determined later. Intuitively, this test plays the same role for \acp{DMC} as the radiometer played for Gaussian channels~\cite{Bash2013} in that it only depends on the codeword weight. To bound the probability of false alarm, we use Theorem~\ref{Berry-Esseen} to obtain 
\begin{align}
\label{eq:fa_first}
\P[H_0]{\sum_{i=1}^n A(Z_i) \geq \tau} & \leq Q\left(\frac{\tau - n \mu_0}{\sqrt{n}\sigma_0}\right) +\frac{6t_0}{\sigma_0^{\modified{3}}\sqrt{n}},
\end{align}
with
\begin{align}
\label{eq:def_moments}
\mu_0 \eqdef \E[Q_0]{A(Z)}, \quad\sigma_0^2 \eqdef \text{Var}_{Q_0}\left(A(Z)\right), \quad t_0 \eqdef \E[Q_0]{|A(Z )-\mu_0|^3}.
\end{align}
Note that all above quantities are finite. For the probability of missed detection, we condition on the codeword transmitted by the channel to obtain \editt{with Theorem~\ref{Berry-Esseen}}
\begin{align}
\P[H_1]{\sum_{i=1}^n A(Z_i) \leq \tau}&= \sum_{m=1}^M\frac{1}{M}\P{\sum_{i=1}^n A(Z_i) \leq \tau |\mathbf{X}=\textbf{x}_m} \\
& \leq \sum_{m=1}^M \frac{1}{M}\left( Q\left(\frac{-\tau + n \mu_0 + \wt{\textbf{x}_m}(\mu_1-\mu_0)}{\sqrt{n\sigma_0^2 + \wt{\textbf{x}_m}\left(\sigma_1^2-\sigma_0^2\right)}}\right)+\frac{6\left(t_0+\wt{\textbf{x}_m}/n(t_1-t_0)\right)}{\left(\sigma_0^2+\wt{\textbf{x}_m}/n(\sigma_1^2-\sigma_0^2)\right)^{3/2}\sqrt{n}}\right)\\
& \leq \sum_{m=1}^M \frac{1}{M}\left( Q\left(\frac{-\tau + n \mu_0 + \wt{\textbf{x}_m}(\mu_1-\mu_0)}{\sqrt{n\sigma_0^2 + \wt{\textbf{x}_m}\left(\sigma_1^2-\sigma_0^2\right)}}\right)\right) + \frac{B_1}{\sqrt{n}},
\end{align}
with
\begin{align}
\mu_1 \eqdef \E[Q_1]{A(Z)}, \quad\sigma_1^2 \eqdef \text{Var}_{Q_1}\left(A(Z)\right), \quad t_1 \eqdef \E[Q_1]{|A(Z )-\mu_1|^3}.
\end{align}
If we choose $\tau = n\mu_0 + \frac{w_{\min}}{2}(\mu_1-\mu_0)$, we have
\begin{align}
\label{eq:md_summation_cw}
\P[H_1]{\sum_{i=1}^n A(Z_i) \leq \tau} \leq \sum_{m=1}^M \frac{1}{M}\left( Q\left(\frac{\wt{\textbf{x}_m}(\mu_1-\mu_0)}{2\sqrt{n\sigma_0^2 + \wt{\textbf{x}_m}\left(\sigma_1^2-\sigma_0^2\right)}}\right)\right) + \frac{B_1}{\sqrt{n}}.
\end{align}
Note that if we have $\sigma_1 \leq\sigma_0$, then we get 
\begin{align}
\beta \eqdef \P[H_1]{\sum_{i=1}^n A(Z_i) \leq \tau} \leq Q\left(\frac{w_{\min} (\mu_1 - \mu_0)}{2\sqrt{n} \sigma_0}\right)+ \frac{B_1}{\sqrt{n}}.
\end{align}
Otherwise \editt{$\sigma_1>\sigma_0$ and} we split the summation in \eqref{eq:md_summation_cw} into two parts: if $\wt{\textbf{x}_m} > \frac{\sigma_1}{\sigma_0} w_{\min}$, then we have
\begin{align}
Q\left(\frac{\wt{\textbf{x}_m}(\mu_1-\mu_0)}{2\sqrt{n\sigma_0^2 + \wt{\textbf{x}_m}\left(\sigma_1^2-\sigma_0^2\right)}}\right) &\stackrel{(a)}{\leq} Q\left(\frac{\frac{\sigma_1}{\sigma_0}w_{\min}(\mu_1 - \mu_0)}{2\sqrt{n}\sigma_1}\right)\\
&= Q\left(\frac{w_{\min} (\mu_1 - \mu_0)}{2\sqrt{n} \sigma_0}\right),
\end{align}
where $(a)$ follows from $\sigma_1 > \sigma_0$. If $w_{\min} \leq \wt{\textbf{x}_m} \leq \frac{\sigma_1}{\sigma_0} w_{\min}$, then we obtain
\begin{align}
Q\left(\frac{\wt{\textbf{x}_m}(\mu_1-\mu_0)}{2\sqrt{n\sigma_0^2 + \wt{\textbf{x}_m}\left(\sigma_1^2-\sigma_0^2\right)}}\right) &\leq Q\left(\frac{w_{\min}(\mu_1-\mu_0)}{2\sqrt{n\sigma_0^2 +\frac{\sigma_1}{\sigma_0} w_{\min}\left(\sigma_1^2-\sigma_0^2\right)}}\right)\\
&=  Q\left(\frac{w_{\min}(\mu_1-\mu_0)}{2\sqrt{n}\sigma_0}\left(\frac{1}{ \sqrt{1+\frac{\sigma_1}{\sigma_0^3}\frac{ w_{\min}}{n}\left(\sigma_1^2-\sigma_0^2\right)}}\right)\right)\\
&\stackrel{(a)}{\leq} Q\left(\frac{w_{\min}(\mu_1-\mu_0)}{2\sqrt{n}\sigma_0}\left(1 - \frac{w_{\min}B_2}{n}\right)\right)\\
&\stackrel{(b)}{\leq} Q\left(\frac{w_{\min}(\mu_1-\mu_0)}{2\sqrt{n}\sigma_0}\right) + \frac{w_{\min}^2 B_3}{n^{\frac{3}{2}}},
\end{align}
for $B_2 =\frac{\sigma_1(\sigma_1^2-\sigma_0^2)}{2\sigma_0^3} > 0$ and $B_3 = \frac{(\mu_1-\mu_0)B_2}{2\sqrt{2\pi}\sigma_0} > 0$ where $(a)$ follows from  $\frac{1}{\sqrt{1 + x}} \geq -\frac{x}{2}+1$ for all $x>0$, and $(b)$ follow from $Q(x-y) \leq Q(x) + \frac{y}{\sqrt{2\pi}}$ for all $0<y<x$. Therefore, we always have
\begin{align}
\label{eq:md_prob}
\beta \leq Q\left(\frac{w_{\min}(\mu_1-\mu_0)}{2\sqrt{n}\sigma_0}\right) + \frac{w_{\min}^2 B_3}{n^{\frac{3}{2}}} + \frac{B_1}{\sqrt{n}}.
\end{align}
Moreover, plugging in the value $\tau$ in \eqref{eq:fa_first}, we obtain
\begin{align}
\label{eq:fa_boudn}
\alpha\eqdef \P[H_0]{\sum_{i=1}^n A(Z_i) \geq \tau} \leq Q\left(\frac{w_{\min}(\mu_1 - \mu_0)}{2\sqrt{n} \sigma_0}\right)+\frac{B_0}{\sqrt{n}},
\end{align}
for some positive constant $B_0$.
Using \eqref{eq:tv_ht}, \eqref{eq:md_prob}, and \eqref{eq:fa_boudn}, we obtain
\begin{align}
\V{\widehat{P}_{\mathbf{Z}}, Q_0^{\pn}} \geq 1 - 2Q\left(\frac{w_{\min}(\mu_1 - \mu_0)}{2\sqrt{n} \sigma_0}\right) - \frac{B_0+B_1}{\sqrt{n}}- \frac{w_{\min}^2B_3}{n^{\frac{3}{2}}}.
\end{align}
Finally, since $\mu_0 = 0$ and $\sigma_0^2  = \mu_1=\chi_2(Q_1\|Q_0)$, we have
\begin{align}
\V{\widehat{P}_\mathbf{Z}, Q_0^{\pn}} \geq 1 - 2Q\left(\frac{w_{\min}\sqrt{\chi_2(Q_1\|Q_0)}}{2\sqrt{n} }\right)- \frac{B_0 + B_1}{\sqrt{n}} - \frac{w_{\min}^2B_3}{n^{\frac{3}{2}}}.
\end{align}
\end{IEEEproof}

Next, we develop a bound for the maximum weight of a sub-codebook.
\begin{lemma}
\label{lm:wt-total-variation}
Let $\mathcal{C}$ be an $(M, K, n, \epsilon, \delta)_V$ code for a binary-input  covert communication channel $(\calX, W_{Y|X}, W_{Z|X}, \calY, \calZ)$. For all $\gamma\in[0,1]$, there exists a subset  of codewords $\mathcal{D}$ such that $|\mathcal{D}|\geq \gamma MK$ and 
\begin{align}
\label{eq:weight-condition}
\frac{1}{\sqrt{n}} \max\{\wt{\textbf{x}}: \textbf{x}\in \mathcal{D}\} \leq \frac{2}{\sqrt{\chi_2(Q_1\|Q_0)}}Q^{-1}\left(\frac{1-\delta}{2}-\frac{C}{\sqrt{n}}-\gamma\right),
\end{align}
where $C$ is a constant that depends only on the channel.
\end{lemma}
\begin{IEEEproof}
We define 
\begin{align}
A \eqdef \frac{2}{\sqrt{\chi_2(Q_1\|Q_0)}}Q^{-1}\left(\frac{1-\delta}{2}-\frac{C_2}{\sqrt{n}}-\gamma\right)
\end{align}
 where $C_2 > 0$ is specified later and $\mathcal{D} \eqdef\{\textbf{x}\in \mathcal{C}: \wt{\textbf{x}} \leq  A\sqrt{n}\}$. \editt{The set} $\mathcal{D}$ satisfies \eqref{eq:weight-condition}, and we just need to check $|\mathcal{D}|\geq \gamma MK$. To this end, let $\widehat{P}_1$ and $\widehat{P}_2$ be the induced output distributions for codes $\mathcal{D}$ and $\mathcal{C}\setminus \mathcal{D}$, respectively. Then, we have
\begin{align}
\delta &\stackrel{(a)}{\geq} \V{\widehat{P}_\mathbf{Z}, Q_0^{\pn}}\\
&\stackrel{(b)}{\geq} (1-\frac{|\mathcal{D}|}{MK}) \V{\widehat{P}_2, Q_0^{\pn}} - \frac{|\mathcal{D}|}{MK}  \V{\widehat{P}_1, Q_0^{\pn}}\\
&\stackrel{(c)}{\geq}(1-\frac{|\mathcal{D}|}{MK})\left(1- \frac{B(1+A^2)}{\sqrt{n}} -2 Q\left(\frac{A\sqrt{n}\sqrt{\chi_2(Q_1\|Q_0)}}{2\sqrt{n} }\right)\right)-\frac{|\mathcal{D}|}{MK}\\
&\geq \left(1- 2Q\left(\frac{A\sqrt{\chi_2(Q_1\|Q_0)}}{2}\right)\right) - \frac{B(1+A^2)}{\sqrt{n}}  - 2\frac{|\mathcal{D}|}{MK}\\
&\stackrel{(d)}{=}\left(1-2\left(\frac{1-\delta}{2}-\frac{C_2}{\sqrt{n}}-\gamma\right)\right)-\frac{B(1+A^2)}{\sqrt{n}}  - 2\frac{|\mathcal{D}|}{MK}\\
&\stackrel{(e)}{\geq}\delta + 2\gamma-2\frac{|\mathcal{D}|}{MK},
\end{align}
where $(a)$ follows from the definition of an $(M, K, n, \epsilon, \delta)_V$ code, $(b)$ follows from the triangle inequality and $\widehat{P}_\mathbf{Z} = \frac{|\mathcal{D}|}{MK} \widehat{P}_1 + (1-\frac{|\mathcal{D}|}{MK})\widehat{P}_2$, $(c)$ follows from Lemma \ref{lm:converse-wmin}, $(d)$ follows from the definition of $A$, and $(e)$ follows by choosing
\begin{align}
C_2 >\frac{1}{2}B\left(1+\left(\frac{2}{\sqrt{\chi_2(Q_1\|Q_0)}}Q^{-1}\left(\frac{1-\delta}{2}-\gamma\right)\right)^2\right).
\end{align}
Therefore, we get $|\mathcal{D}|\geq \gamma MK$.
\end{IEEEproof}

We can now establish the converse.
\begin{IEEEproof}[\typo{Converse for Theorem~\ref{th:main_result_tv}}]
\editt{
In Lemma~\ref{lm:wt-total-variation}, if we set $\gamma = \frac{1}{\sqrt{n}}$, for any $(M, K, n, \epsilon, \delta)_V$ code, we \editt{obtain} a subset of codewords $\mathcal{D}$ with $|\mathcal{D}| \geq MK/\sqrt{n}$ and 
\begin{align}
\max_{\textbf{x} \in \mathcal{D}} \frac{\wt{\textbf{x}}}{\sqrt{n}} &\leq \frac{2}{\sqrt{\chi_2(Q_1\|Q_0)}}Q^{-1}\left(\frac{1-\delta}{2}-\frac{C}{\sqrt{n}}-\gamma\right)\\
&= \frac{2}{\sqrt{\chi_2(Q_1\|Q_0)}}Q^{-1}\left(\frac{1-\delta}{2}\right)+O\left(\frac{1}{\sqrt{n}}\right).
\end{align}
Therefore, by Theorem~\ref{thm:converse-general}, we have
\begin{align}
\log M^*_V(n, \epsilon, \delta) \leq \frac{2}{\sqrt{\chi_2(Q_1\|Q_0)}}Q^{-1}\left(\frac{1-\delta}{2}\right)D_Pn^{\frac{1}{2}}-\sqrt{\frac{2}{\sqrt{\chi_2(Q_1\|Q_0)}}Q^{-1}\left(\frac{1-\delta}{2}\right)V_P}Q^{-1}(\epsilon)n^{\frac{1}{4}} + O(\log n).
\end{align}}

%

  \editt{Lemma~\ref{lm:wt-total-variation} already ensures the upper bound needed to apply Theorem~\ref{thm:converse-general}, and we only need to make sure that we can choose $\gamma$ in Lemma~\ref{lm:wt-total-variation} such that we can satisfy the other conditions in~\eqref{eq:wt-condition-metric-first-order}. For every $(M_n,K_n,n,\epsilon_n,\delta)_V$ code in a sequence of codes such that $\lim_{n\to\infty}\epsilon_n=0$, we choose $\gamma_n=\frac{1}{n}$. Then,
    \begin{align}
     \lim_{n\to\infty}-\frac{\log \gamma_n}{\sqrt{n}}=0.  
    \end{align}
  Hence, the results follows by Theorem~\ref{thm:converse-general}.} 

\end{IEEEproof}

\subsection{Covertness with probability of missed detection at fixed significance level}
\label{sec:covert-comm-missed-detection}
In this subsection, we prove Theorem~\ref{th:main_result_beta}.
\begin{lemma}
For a fixed $\alpha\in]0,1[$ and two distributions $P$ and $Q$ over same set $\mathcal{Z}$, we define $d(P, Q) \eqdef 1 - \alpha - \beta_\alpha(Q, P)$; then $d$ is a quasi-metric, and we have
\begin{align}
d(R, Q) \leq d(P, Q) + \D{R}{P}+\sqrt{\D{R}{P}}\max\left(1,\log \frac{1}{\min_z Q(z)}\right).
\end{align}
\end{lemma}
\begin{IEEEproof}
By definition of $\beta_\alpha(Q,R)$, there exists a set $\mathcal{T}$ with $Q(\mathcal{T}) \leq \alpha$ and $R(\mathcal{T}) = \beta_\alpha(Q, R)$; therefore we have
\begin{align}
\beta_\alpha(Q, P) &\leq P(\mathcal{T})\\
&=R(\mathcal{T}) + (P(\mathcal{T}) - R(\mathcal{T})) \displaybreak[0]\\
&\leq R(\mathcal{T}) + \V{R, P}\\
&= \beta_\alpha(Q, R) + \V{R, P}.
\end{align}
By Pinsker's Inequality, we obtain
\begin{align}
\beta_\alpha(Q, R) + \V{R, P} &\leq \beta_\alpha(Q, R) + \sqrt{\D{R}{P}}.
\end{align}
Finally, using the definition of $d(P, Q)$, we get the result.
\end{IEEEproof}

\begin{IEEEproof}[\typo{Achievability of Theorem~\ref{th:main_result_beta}}] \modified{Fix $\epsilon\in]0, 1[$, $\delta \in ]0, 1- \alpha[$, and $n$; we just show how we can choose $\ell_n$ such that $\beta_{\alpha}(Q_0^\pn, \ppmZ{n}{\ell_n}) \geq 1 - \alpha - \delta + \frac{1}{\sqrt{n}}$ and the rest of the proof is similar to the proof for variational distance. Let $\Lambda \eqdef Q^{-1}(1-\alpha-\delta)$ and $\Upsilon \eqdef Q^{-1}(\alpha)$; we \editt{wish to choose} $\ell_n$ to satisfy
\begin{align}
\label{eq:ell_beta}
\ell_n \sqrt{\frac{\chi_2(Q_1\|Q_0)}{n} } - Q^{-1}\pr{\alpha + \frac{1}{\sqrt{\ell_n}}} \leq \Lambda - \frac{\sqrt{2\pi} e^{\frac{\Lambda^2}{2}}}{\sqrt{\ell_n}}  + O\pr{\frac{1}{\sqrt{n}}}
\end{align}
Using $Q^{-1}\pr{\alpha + x} = Q^{-1}(\alpha)  - \sqrt{2\pi} e^{\frac{\pr{Q^{-1}(\alpha)}^2}{2}}x + O(x^2) = \Upsilon +\sqrt{2\pi} e^{\frac{\Upsilon^2}{2}}x + O(x^2)$ for $x$ close to zero, the above inequality is equivalent to
\begin{align}
\ell_n \sqrt{\frac{\chi_2(Q_1\|Q_0)}{n} } - \Upsilon + \frac{\sqrt{2\pi} e^{\frac{\Upsilon^2}{2}}}{\sqrt{\ell_n}} \leq\Lambda - \frac{\sqrt{2\pi} e^{\frac{\Lambda^2}{2}}}{\sqrt{\ell_n}}  + O\pr{\frac{1}{\sqrt{n}}}.
\end{align}
We form \editt{the} cubic equation
\begin{align}
x^3  -\pr{\Lambda + \Upsilon + O\pr{\frac{1}{\sqrt{n}}}}\sqrt{\frac{n}{\chi_2(Q_1\|Q_0)}}x + \sqrt{2\pi}\pr{e^{\frac{\Lambda^2}{2}} + e^{\frac{\Upsilon^2}{2}}}\sqrt{\frac{n}{\chi_2(Q_1\|Q_0)}}=0,
\end{align}
and set $\sqrt{\ell_n}$ to be its solution; Thus, \editt{similar} to~\eqref{eq:sq_ell_v},  we obtain
\begin{align}
\sqrt{\ell_n} 
&= 2\sqrt{\frac{\pr{\Lambda + \Upsilon + O\pr{\frac{1}{\sqrt{n}}}} \sqrt{n}}{3\sqrt{\chi_2(Q_1\|Q_0)}}}\cos\pr{\frac{1}{3}\arccos\pr{\frac{3\sqrt{2\pi} \pr{e^{\frac{\Lambda^2}{2}}+ e^{\frac{\Upsilon^2}{2}}}}{-2\pr{\Lambda + \Upsilon + O\pr{\frac{1}{\sqrt{n}}}}}\sqrt{\frac{3\sqrt{\chi_2(Q_1\|Q_0)}}{\pr{\Lambda + \Upsilon + O\pr{\frac{1}{\sqrt{n}}}} \sqrt{n}}} }}\\
&= 2\sqrt{\frac{\pr{\Lambda + \Upsilon + O\pr{\frac{1}{\sqrt{n}}}} \sqrt{n}}{3\sqrt{\chi_2(Q_1\|Q_0)}}}\cos\pr{\frac{1}{3}\pr{\frac{\pi}{2}+\frac{3\sqrt{2\pi} \pr{e^{\frac{\Lambda^2}{2}}+ e^{\frac{\Upsilon^2}{2}}}}{2\pr{\Lambda + \Upsilon }}\sqrt{\frac{3\sqrt{\chi_2(Q_1\|Q_0)}}{\pr{\Lambda + \Upsilon } \sqrt{n}}} + O\pr{\frac{1}{\sqrt{n}}} }}\\
&= 2\sqrt{\frac{\pr{\Lambda + \Upsilon + O\pr{\frac{1}{\sqrt{n}}}} \sqrt{n}}{3\sqrt{\chi_2(Q_1\|Q_0)}}}\pr{\cos\pr{\frac{\pi}{6}}-\sin\pr{\frac{\pi}{6}}\frac{\sqrt{2\pi} \pr{e^{\frac{\Lambda^2}{2}}+ e^{\frac{\Upsilon^2}{2}}}}{2\pr{\Lambda + \Upsilon }}\sqrt{\frac{3\sqrt{\chi_2(Q_1\|Q_0)}}{\pr{\Lambda + \Upsilon } \sqrt{n}}} + O\pr{\frac{1}{\sqrt{n}}} }\\
&= \sqrt{\frac{(\Lambda + \Upsilon) \sqrt{n}}{\sqrt{\chi_2(Q_1\|Q_0)}}} - \sqrt{\frac{\pi}{2}} \frac{e^{\frac{\Lambda^2}{2}} + e^{\frac{\Upsilon^2}{2}}}{\Lambda + \Upsilon} + O\pr{\frac{1}{\sqrt{n}}}.
\end{align}
Hence, if we choose
\begin{align}
\ell_n = \frac{(\Lambda + \Upsilon)}{\sqrt{\chi_2(Q_1\|Q_0)}}n^{\frac{1}{2}} - \frac{\sqrt{2\pi}\pr{e^{\frac{\Lambda^2}{2}} + e^{\frac{\Upsilon^2}{2}}}}{\sqrt{\Lambda + \Upsilon} \chi_2(Q_1\|Q_0) ^{\frac{1}{4}}} n^{\frac{1}{4}}  +O(1),
\end{align}
 we have
\begin{align}
\beta_{\alpha}(Q_0^\pn, \ppmZ{n}{\ell_n}) 
&\stackrel{(a)}{\geq} Q\pr{\ell_n\sqrt{\frac{\chi_2(Q_1\|Q_0)}{n}} - Q^{-1}\pr{\alpha + \frac{1}{\ell_n}}} - \frac{1}{\sqrt{\ell_n}} + O\pr{\frac{1}{\sqrt{n}}}\\
&\stackrel{(b)}{\geq} Q\pr{\Lambda - \frac{\sqrt{2\pi} e^{\frac{\Lambda^2}{2}}}{\sqrt{\ell_n}}  + O\pr{\frac{1}{\sqrt{n}}}} - \frac{1}{\sqrt{\ell_n}} + O\pr{\frac{1}{\sqrt{n}}}\\
&\stackrel{(c)}{\geq} Q\pr{\Lambda} + \frac{1}{\sqrt{\ell_n}} - \frac{1}{\sqrt{\ell_n}} + O\pr{\frac{1}{\sqrt{n}}}\\
&= 1 - \alpha - \delta + O\pr{\frac{1}{\sqrt{n}}},
\end{align}
where $(a)$ follows from \eqref{eq:beta_ppm}, $(b)$ follows from \eqref{eq:ell_beta}, and $(c)$ follows from $Q(\Lambda + x) = Q(\Lambda) + \frac{x}{\sqrt{2\pi} e^{\frac{\Lambda^2}{2}}} + O(x^2)$ for $x$ close to zero. }
\end{IEEEproof}

The proof of the converse of Theorem~\ref{th:main_result_beta} requires the  following steps similar to those of covertness with variational distance. We first relate the probability of missed detection $\beta_\alpha\left(Q_0^{\pn}, \widehat{P}_\mathbf{Z}\right)$ at significance level $\alpha$  to the minimum weight of codewords.
\begin{lemma}
\label{lm:converse-wmin-beta}
Consider a binary-input \ac{DMC} $(\calX, W_{Z|X}, \calZ)$ and $M$ codewords $\textbf{x}_1, \cdots, \textbf{x}_M\in\mathcal{X}^n$ with induced distribution $\widehat{P}_\mathbf{Z}$ on $\mathcal{Z}^n$. If $\frac{\sqrt{n}Q^{-1}(\alpha)}{\sqrt{\chi_2(Q_1\|Q_0)}}<w_{\min} \eqdef \min_{m\in \intseq{1}{M}}\wt{\textbf{x}_m}$,  we have
\begin{align}
\beta_\alpha\left(Q_0^{\pn}, \widehat{P}_\mathbf{Z}\right) \leq Q\left(\frac{w_{\min}\sqrt{\chi_2(Q_1\|Q_0)}}{\sqrt{n}}-Q^{-1}(\alpha)\right)+ \frac{B}{\sqrt{n}} + \frac{w_{\min}^2B}{n^{\frac{3}{2}}},
\end{align}
where $B$ is a constant that \editt{just depends} on the channel.
\end{lemma}

\begin{IEEEproof}
For the hypothesis testing problem consisting of two hypotheses $H_0$ with distribution $Q_0^\pn$ and $H_1$ with distribution $\widehat{P}_\mathbf{Z}$, we introduce again the test
\begin{align}
T(\textbf{z}) \eqdef \indic{\sum_{i=1}^n A(z_i) >\tau},
\end{align}
where $A(z) \eqdef \frac{Q_1(z)-Q_0(z)}{Q_0(z)}$. Using Theorem~\ref{Berry-Esseen} and the calculations in the proof of Lemma~\ref{lm:converse-wmin}, we have
\begin{align}
\P[H_0]{\sum_{i=1}^n A(Z_i) \geq \tau} &= Q\left(\frac{\tau}{\sqrt{n\chi_2(Q_1\|Q_0)}}\right)+ O\left(\frac{1}{\sqrt{n}}\right).
\end{align}
Thus, if we choose $\tau = \sqrt{n\chi_2(Q_1\|Q_0)}Q^{-1}\left(\alpha+O\left(n^{-\frac{1}{2}}\right)\right)$, the false alarm probability would be less than or equal to $\alpha$. Hence, by definition of $\beta_\alpha\left(Q_0^{\pn}, \widehat{P}_\mathbf{Z}\right)$, we get
\begin{align}
\beta_\alpha\left(Q_0^{\pn}, \widehat{P}_\mathbf{Z}\right) &\leq \P[H_1]{\sum_{i=1}^n A(Z_i) \leq\tau }\\
&= \sum_{i=1}^M\frac{1}{M}\P{\sum_{i=1}^n A(Z_i) \leq \tau |\mathbf{X}=\textbf{x}_i}\\
&= \sum_{i=1}^M \frac{1}{M}\left( Q\left(\frac{-\tau + \chi_2(Q_1\|Q_0)\wt{\textbf{x}_i}}{\sqrt{n\chi_2(Q_1\|Q_0)+ \wt{\textbf{x}_i}(\sigma_1^1 - \sigma_0^2)}} \right)+O\left(n^{-\frac{1}{2}}\right)\right),
\end{align}
with
\begin{align}
\sigma_0^2 \eqdef\text{Var}_{Q_0}(A(Z)) \text{ and } \sigma_1^2\eqdef\text{Var}_{Q_1}(A(Z)).
\end{align}
Plugging in the value of $\tau$, we obtain
\begin{align}
\beta_\alpha\left(Q_0^{\pn}, \widehat{P}_\mathbf{Z}\right) &\leq \sum_{i=1}^M \frac{1}{M}\left( Q\left(\frac{-\sqrt{n\chi_2(Q_1\|Q_0)}Q^{-1}\left(\alpha+O\left(n^{-\frac{1}{2}}\right)\right) + \chi_2(Q_1\|Q_0)\wt{\textbf{x}_i}}{\sqrt{n\chi_2(Q_1\|Q_0)+ \wt{\textbf{x}_i}(\sigma_1^1 - \sigma_0^2)}} \right)+O\left(n^{-\frac{1}{2}}\right)\right)\\
&= \sum_{i=1}^M \frac{1}{M}\left( Q\left(\frac{-Q^{-1}\left(\alpha\right)}{\sqrt{1+ \frac{\wt{\textbf{x}_i}}{n}\frac{\sigma_1^2- \sigma_0^2}{\chi_2(Q_1\|Q_0)}}} + \frac{\chi_2(Q_1\|Q_0)\wt{\textbf{x}_i}}{\sqrt{n\chi_2(Q_1\|Q_0)+ \wt{\textbf{x}_i}(\sigma_1^1 - \sigma_0^2)}} \right)+O\left(n^{-\frac{1}{2}}\right)\right).
\end{align}
Analogous to the proof of Lemma~\ref{lm:converse-wmin}, we obtain
\begin{align}
\beta_\alpha\left(Q_0^{\pn}, \widehat{P}_\mathbf{Z}\right) &\leq  Q\left(\frac{w_{\min}\sqrt{\chi_2(Q_1\|Q_0)}}{\sqrt{n}}-Q^{-1}(\alpha)\right)+O\left(n^{-\frac{1}{2}}\right) + O\left(\frac{w_{\min}^2}{n^{\frac{3}{2}}}\right).
\end{align}
\end{IEEEproof}
Next, we develop an upper bound on the weight of codewords as a function of channel characteristics and covertness measured in terms of probability of missed detection. 
\begin{lemma}
\label{lm:converse_beta_max_wt}
Let $(\calX, W_{Y|X}, W_{Z|X}, \calY, \calZ)$ be a binary-input covert communication channel and $\mathcal{C}$ be an $(M, K, n, \epsilon, \delta, \alpha)_{\beta}$ code. For all $\gamma\in[0,1]$, we can choose a subset  of codewords $\mathcal{D}$ such that $|\mathcal{D}|\geq \gamma MK$ and 
\begin{align}
\label{eq:weight-condition-beta}
\frac{1}{\sqrt{n}} \max\{\wt{\textbf{x}}: \textbf{x}\in \mathcal{D}\} \leq \frac{1}{\sqrt{\chi_2(Q_1\|Q_0)}}\left(Q^{-1}\left(1-\alpha-\delta-\frac{C}{\sqrt{n}}-\gamma\right) + Q^{-1}(\alpha)\right),
\end{align}
where $C$ is a constant that depends just on the channel.
\end{lemma}

\begin{IEEEproof}
If $A \eqdef \frac{1}{\sqrt{\chi_2(Q_1\|Q_0)}}\left(Q^{-1}\left(1-\alpha-\delta-\frac{C_2}{\sqrt{n}}-\gamma\right) + Q^{-1}(\alpha)\right)$ for $C_2>0$ specified later, we define
\begin{align}
\mathcal{D} \eqdef\{\textbf{x}\in \mathcal{C}: \wt{\textbf{x}} \leq  A\sqrt{n}\}.
\end{align}
Obviously, $\mathcal{D}$ satisfies \eqref{eq:weight-condition-beta}, and we just need to check $|\mathcal{D}|\geq \gamma MK$. To do so, let $\widehat{P}_1$ and $\widehat{P}_2$ be the induced output distributions for codes $\mathcal{D}$ and $\mathcal{C}\setminus \mathcal{D}$, respectively. Using Lemma~\ref{lm:converse-wmin-beta} for $\mathcal{C}\setminus\mathcal{D}$, we get
\begin{align}
\beta_\alpha\left(Q_0^\pn, \widehat{P}_2\right) &\leq  Q\left(A\sqrt{\chi_2(Q_1\|Q_0)}-Q^{-1}(\alpha)\right)+\frac{B(1+A^2)}{\sqrt{n}}\\
&\stackrel{(a)}{\leq} 1-\alpha-\delta -\gamma,
\end{align}
where $(a)$ follows from the definition of $A$ and by choosing
\begin{align}
C_2 > \frac{1}{2}B\left(1+\left(\frac{1}{\sqrt{\chi_2(Q_1\|Q_0)}}Q^{-1}\left(1-\alpha-\delta-\gamma\right)\right)^2\right).
\end{align}
This means that there exists $\mathcal{T} \subset \mathcal{Z}^n$ such that $Q_0^\pn(\mathcal{T}) \leq \alpha$ and $\widehat{P}_2(\mathcal{T}) \leq 1-\alpha-\delta-\gamma$. Accordingly, we have
\begin{align}
1-\alpha-\delta&\stackrel{(a)}{\leq} \beta_\alpha\left(Q_0^\pn, \widehat{P}_\mathbf{Z}\right) \\
&\leq \widehat{P}_\mathbf{Z}(\mathcal{T}) \\
&\stackrel{(b)}{=}\frac{|\mathcal{D}|}{MK} \widehat{P}_1(\mathcal{T}) + (1-\frac{|\mathcal{D}|}{MK})\widehat{P}_2(\mathcal{T})\\
&\leq \frac{|\mathcal{D}|}{MK} +  1-\alpha-\delta-\gamma,
\end{align}
where $(a)$ follows from the definition of an $(M, K, n, \epsilon, \delta, \alpha)_{\beta}$ code, and $(b)$ follows from $\widehat{P}_{\mathbf{Z}} = \frac{|\mathcal{D}|}{MK} \widehat{P}_1 + (1-\frac{|\mathcal{D}|}{MK})\widehat{P}_2$. Simplifying the above inequality, we get $|\mathcal{D}|\geq \gamma MK$.
\end{IEEEproof}
\editt{The converse then proceeds as for the converse of Theorem 2 in the previous section.}

\section{Conclusion}
\label{sec:conclusion}
We have developed \editt{an approach} to study covert communication when covertness is measured with several ``quasi-metrics,'' as defined in Section~\ref{sec:covert-comm-results}. It is legitimate to ask which metric would make most sense from an operational perspective. While relative entropy is amenable to a fairly extensive information-theoretic analysis, as illustrated by our complete characterization of second-order asymptotics in Theorem~\ref{th:main_result_d}, variational distance and probability of false alarm are probably more adequate since they directly relate to the operation of an adversary attempting to detect communication. Since measuring covertness in terms of variational distance does not impose any constraint on where the adversary operates on its \ac{ROC} curve, we believe that variational distance is perhaps the \editt{most operationally relevant} covertness metric.

Our results \editt{have been restricted} to binary-input~\acp{DMC} \editt{to obtain simple closed-form expressions} but extensions to arbitrary finite input alphabets \editt{may be obtained following} the approach of~\cite{Wang2016b}; \todo{we'll have to double check this carefully. the achievability is fine, not sure about the converse} however several research questions remain open. We have not characterized the exact second-order asymptotics of covert communication with variational distance and probability of missed detection metrics; we conjecture, however, that the upper-bounds are achievable. We have also not characterized the second-order asymptotics for the number of key bits required; a close inspection of our current proof technique shows that we explicitly rely on a law of large numbers to analyze the number of key bits, which one would have to circumvent.

\todo{perhaps comment on \ac{PPM}}

\appendices

\section{The Effect of Average Probability of Error on First Order Asymptotics}
\label{sec:effect-aver-prob}
\modified{Assume that the optimal second order asymptotics are of the form
\begin{align}
  \label{eq:assumption_scaling}
    \log M^*(\epsilon,\delta) = f(\delta)\sqrt{n} + o(\sqrt{n})
  \end{align}
  for some function $f$ that is \emph{strictly} increasing in $\delta$ and does not depend on $\epsilon$, which is what we \editt{would hope} to establish \editt{based on our results maximum probability of error}. Now for $1>\epsilon>0$ and $\delta>0$, pick $\frac{\epsilon}{1-\epsilon}>\alpha>0$ and set $\epsilon'=(1+\alpha)\epsilon-\alpha$, and $\delta'=(1+\alpha)\delta$. Since $\delta'>\delta$ and because of~\eqref{eq:assumption_scaling}, we expect for all blocklengths large enough that 
  \begin{align*}
\log M^*(\epsilon',\delta') >     \log M^*(\epsilon,\delta).
  \end{align*}
Consider now a code with $M^*(\epsilon',\delta')$ codewords, and let $\widehat{Q'}^n$ be the distribution induced at the eavesdropper's output when randomizing uniformly over all codewords. \editt{Let $\mathcal{S}$ be a set (with repetition) of $N\eqdef \alpha M^*(\epsilon',\delta')$ all-zero codewords.} We construct a new code by adding the set $\mathcal{S}$ to the previous code. The resulting code has average probability of error 
  \begin{align*}
P_{\text{avg}} &= \P{\widehat{W}\neq W|W\in\mathcal{S}} \P{W\in\mathcal{S}}+ \P{\widehat{W}\neq W|W\notin\mathcal{S}} \P{W\notin\mathcal{S}}\\
    &\leq \frac{M^*(\epsilon',\delta')}{M^*(\epsilon',\delta')+N}\epsilon' +\frac{N}{M^*(\epsilon',\delta')+N}=\frac{1}{1+\alpha}\epsilon'+\frac{\alpha}{1+\alpha}=\epsilon.
  \end{align*}
It also induces a new distribution $\widehat{Q}^n$ at the eavesdropper's output when randomizing uniformly over all codewords, such that 
  \begin{align*}
\avgD{\widehat{Q'}^n}{Q_0^{\otimes n}} &= \avgD{\frac{M^*(\epsilon',\delta')}{M^*(\epsilon',\delta')+N}\widehat{Q}^n+\frac{N}{M^*(\epsilon',\delta')+N}Q_0^{\otimes n}}{Q_0^{\otimes n}}\\
      &\leq \frac{M^*(\epsilon',\delta')}{M^*(\epsilon',\delta')+N}\delta' = \frac{1}{1+\alpha}\delta'=\delta.
  \end{align*}
  We therefore obtain a new code with probability of error less than $\epsilon$ and relative entropy less than $\delta$, but with a number of codewords $M^*(\epsilon',\delta') + \editt{N}$ that \emph{exceeds} $ M^*(\epsilon,\delta)$. This contradiction shows that~\eqref{eq:assumption_scaling} does not hold, so that either $f(\delta)$ is non-increasing or $\delta$ or depends on $\epsilon$. In both cases, this suggests a result significantly more complex that what we were trying to establish. Note that this does not contradict existing results on covert capacity~\cite{Wang2016b,Bloch2016a} because these works have all been in the regime $\lim_{n\rightarrow\infty}\epsilon=0$. \editt{ Finally, from this example, we conclude that no strong converse exists when measuring reliability with an \emph{average} probability of error and even the first-order asymptotics \editt{depend} on a non-vanishing average probability of error. Hence, one could expect a much more challenging analysis of second-order asymptotics under average probability of error constraint.}}

\section{A technical lemma}
\label{sec:technical-lemma}

\begin{lemma}
\label{lm:A-moments}
\editt{Suppose $\mathbf{Z} \eqdef (Z_1, \cdots, Z_m)$ is distributed according to $Q_0^{\proddist m}$. Define $A(z) \eqdef \frac{Q_1(z) - Q_0(z)}{Q_0(z)}$ and let $B\eqdef\frac{1}{m}\sum_{i=1}^m A(Z_i)$. Also define $\underline{B}$ with distribution $\P{\underline{B}=b}\eqdef \P{B=b|B\neq -1}$ and $\underline{C} \eqdef \log(1+\underline{B})$. There exists a constant $0\leq \tau<1$ depending only on the channel such that
\begin{align}
\E{\underline{B}} &= O(\tau^m),\\
\E{\underline{B}^2} &= \frac{\chi_2(Q_1\|Q_0)}{m} + O(\tau^m),\\
\E{\underline{B}^3} &= O\pr{\frac{1}{m^2}},\\
\E{\underline{B}^4} &= \frac{\chi_2(Q_1\|Q_0)^2}{m^2} + O\pr{\frac{1}{m^3}},
\end{align}
and
\begin{align}
\E{\underline{C}} &= -\frac{\chi_2(Q_1\|Q_0)}{2m} + O\pr{\frac{1}{m^2}}\\
\Var{\underline{C}} &= \frac{\chi_2(Q_1\|Q_0)}{m}+ O\pr{\frac{1}{m^2}}\\ 
\E{|\underline{C}-\E{\underline{C}}|^3} &\leq\frac{\chi_2(Q_1\|Q_0)^{\frac{3}{2}}}{m^{\frac{3}{2}}} + O\pr{\frac{1}{m^{\frac{9}{4}}}}.
\end{align}
Suppose $\widetilde{\mathbf{Z}} \eqdef (\widetilde{Z}_1, \cdots, \widetilde{Z}_m)$ is distributed according to $\ppmZ{m}{1}$. Define $\widetilde{B} \eqdef \frac{1}{m} \sum_{i=1}^m A(\widetilde{Z}_i)$, $\underline{\widetilde{B}}$ with distribution $\P{\widetilde{\underline{B}} = b} \eqdef \P{\widetilde{B} = b| \widetilde{B} \neq -1}$ and $\widetilde{\underline{C}} \eqdef \log(1+\underline{\widetilde{B}})$. Then
\begin{align}
\E{\widetilde{\underline{B}}} &= \frac{\chi_2(Q_1\|Q_0)}{m} +  O(\tau^m), \\
\E{\widetilde{\underline{B}}^2} &= \frac{\chi_2(Q_1\|Q_0)}{m }+ O\pr{\frac{1}{m^2}},\\
\E{\widetilde{\underline{B}}^3} &= O\pr{\frac{1}{m^2}},\\
\E{\widetilde{\underline{B}}^4} &= \frac{\chi_2(Q_1\|Q_0)^2}{m^2} + O\pr{\frac{1}{m^3}},
\end{align}
and
\begin{align}
\E{\widetilde{\underline{C}}} &= \frac{\chi_2(Q_1\|Q_0)}{2m} + O\pr{\frac{1}{m^2}}\\
\Var{\widetilde{\underline{C}}} &= \frac{\chi_2(Q_1\|Q_0)}{m}+ O\pr{\frac{1}{m^2}}\\ 
\E{|\widetilde{\underline{C}}-\E{\widetilde{\underline{C}}}|^3} &\leq \frac{\chi_2(Q_1\|Q_0)^{\frac{3}{2}}}{m^{\frac{3}{2}}} + O\pr{\frac{1}{m^{\frac{9}{4}}}}.
\end{align}}
\end{lemma}
\editt{
\begin{IEEEproof}
We first show that the moments of $B$ and $\underline{B}$ are close. Define $\kappa \eqdef\max_{z: Q_0(z) > 0} \abs{\frac{Q_1(z)-Q_0(z)}{Q_0(z)}}$ and $\tau \eqdef \P[Q_0]{A(Z) = -1}$. Notice that $\V{P_{B}, P_{\underline{B}}} = \tau^m$, $\abs{B}\leq\kappa$ and $\abs{\underline{B}}\leq \kappa$ almost surely and that $\kappa$ does not depend on $m$. For any $q \geq 1$, we therefore obtain 
\begin{align}
\abs{\E{B^q} - \E{\underline{B}^q}}
&\leq \sum_b \abs{b^q\pr{ \P{B = b} - \P{\underline{B} = b}}}\\
&\leq \kappa^q \V{P_{B}, P_{\underline{B}}}\\
&=  \kappa^q \tau^m.
\end{align}
Therefore, it is sufficient to find $\E{B^i}$ for $i=1, 2, 3, 4$. Note that 
\begin{align}
\E[Q_0]{A(Z)} &= 0\\
\E[Q_1]{ A(Z)} &= \chi_2(Q_1\|Q_0)\\
\text{Var}_{Q_0} (A(Z)) &= \chi_2(Q_1\|Q_0).
\end{align}
Hence, we obtain
\begin{align}
\E{{B}} &= \E{\frac{1}{m} \sum_{i=1}^m A(Z_i)} = 0,\displaybreak[0]\\
\E{{B}^2} &= \frac{1}{m^2}\E{\pr{\sum_{i=1}^m A(Z_i)}^2} = \frac{1}{m^2} \sum_{i=1}^m\sum_{j=1}^m \E{A(Z_i) A(Z_j)} = \frac{\chi_2(Q_1\|Q_0)}{m}\displaybreak[0]\\
\E{B^3} &= \frac{1}{m^3}\E{\sum_{i=1}^m\sum_{j=1}^m \sum_{k=1}^m A(Z_i)A(Z_k) A(Z_j)} = \frac{\sum_{i=1}^m\E{A(Z_i)^3}}{m^3} = O\pr{\frac{1}{m^2}}\displaybreak[0]\\
  \E{B^4} &= \frac{1}{m^4}\E{\sum_{i=1}^m\sum_{j=1}^m\sum_{k=1}^m\sum_{t=1}^m A(Z_i)A(Z_j)A(Z_k)A(Z_t)}\\
  &=\frac{\sum_{i=1}^m A(Z_i)^4 + 2\sum_{i=1}^m \sum_{j=i+1}^m \E{A(Z_i)^2}\E{A(Z_j)^2}}{m^4}\displaybreak[0]\\
&=\frac{\chi_2(Q_1\|Q_0)^2}{m^2} +\frac{-\chi_2(Q_1\|Q_0) + \E[Q_0]{A(Z)^4}}{m^3} = \frac{\chi_2(Q_1\|Q_0)^2}{m^2} + O\pr{\frac{1}{m^3}}.
\end{align}
To find the expected value of $\underline{C}$, note that $ \ln(1+x) \leq x - \frac{x^2}{2} + \frac{x^3}{3}$, and therefore, we have
\begin{align}
\E{\underline{C}} 
&= \E{\log (1+\underline{B})} \leq \E{\underline{B}}- \E{\frac{\underline{B}^2}{2}} + \E{\frac{\underline{B}^3}{3}} = -\frac{\chi_2(Q_1\|Q_0)}{2m} + O\pr{\frac{1}{m^2}}.
\end{align}
For any $0<a<1$ and $x > -a$, we also have $x - \frac{x^2}{2} +\frac{x^3}{3} -\frac{x^4}{4(1-a)^4} \leq \ln(1+x)$; thus,
\begin{align}
\E{\underline{C}} 
&\geq \E{\underline{B}} - \E{\frac{\underline{B}^2}{2}} + \E{\frac{\underline{B}^3}{3}} - \E{\frac{\underline{B}^4}{4(1+\min_{b:\P{\underline{B}=b} > 0}b)^4}}\\
&= -\frac{\chi_2(Q_1\|Q_0)}{2m} + O\pr{\frac{1}{m^2}}  -   \frac{\frac{\chi_2(Q_1\|Q_0)^2}{m^2} + O\pr{\frac{1}{m^3}}}{4(1+\min_{b:\P{\underline{B}=b} > 0}b)^4}\\
&= -\frac{\chi_2(Q_1\|Q_0)}{2m} + O\pr{\frac{1}{m^2}}  -   \frac{\frac{\chi_2(Q_1\|Q_0)^2}{m^2} + O\pr{\frac{1}{m^3}}}{4(1+\min_{z:Q_0(z) > 0, Q_1(z)}A(z))^4}\\
&\stackrel{(a)}{=}  -\frac{\chi_2(Q_1\|Q_0)}{2m}  + O\pr{\frac{1}{m^2}},
\end{align}
where $(a)$ follows since  $\min_{z:Q_0(z) > 0, Q_1(z) > 0} A(z) > -1$. Furthermore, for the variance of $\underline{C}$, since $\log^2(1+x) \geq \pr{x - \frac{x^2}{2}}^2 - x^4$, we have
\begin{align}
\Var{\underline{C}} 
&= \E{\underline{C}^2} - \E{\underline{C}}^2 \\
&\geq \E{\pr{\underline{B} - \frac{\underline{B}^2}{2}}^2} - \E{\underline{B}^4} + O\pr{\frac{1}{m^2}}\\
&= \E{\underline{B}^2} -\E{\underline{B}^3} - \E{\frac{3}{4}\underline{B}^4} + O\pr{\frac{1}{m^2}}\\
&= \frac{\chi_2(Q_1\|Q_0)}{m} + O\pr{\frac{1}{m^2}}
\end{align} 
Moreover, for $-1 < a < 0$ and $x > -a$, we have $\log^2(1+x) \leq \pr{x-\frac{x^2}{2} + \frac{x^3}{3(1-a)^3}}^2$. Therefore,
\begin{align}
\Var{\underline{C}} 
&= \E{\underline{C}^2} - \E{\underline{C}}^2 \\
&\geq \E{\pr{\underline{B}-\frac{\underline{B}^2}{2} - \frac{\underline{B}^3}{3(1-\min_{z:Q_0(z) > 0, Q_1(z) > 0}A(z))^3}}} + O\pr{\frac{1}{m^2}}\\
&= \frac{\chi_2(Q_1\|Q_0)}{m} + O\pr{\frac{1}{m^2}}.
\end{align}
Finally, for the third moment of $C$, note that
\begin{align}
\E{|\underline{C} - \E{\underline{C}}|^3}
&= \E{\pr{\pr{\underline{C}-\E{\underline{C}}}^4}^{\frac{3}{4}}}\\
&\stackrel{(a)}{\leq} \pr{\E{\pr{\underline{C}-\E{\underline{C}}}^4}}^{\frac{3}{4}}\\
&=  \pr{\E{\underline{C}^4 - 4\underline{C}^3 \E{\underline{C}} + 6\underline{C}^2 (\E{\underline{C}})^2 - 4 C (\E{\underline{C}})^3 + (\E{\underline{C}})^4 }}^{\frac{3}{4}}\\
&= \frac{\chi_2(Q_1\|Q_0)^{\frac{3}{2}}}{m^{\frac{3}{2}}} + O\pr{\frac{1}{m^{\frac{9}{4}}}},
\end{align}
where $(a)$ follows from Jensen's inequality.

We now calculate the moments with respect to \ac{PPM} distribution. First note that if we define $P_{\mathbf{Z}}^i(\mathbf{z}) \eqdef Q_1(z_i) \prod_{j\neq i} Q_0(z_j)$, then we can write $\ppmZ{m}{1} = \frac{1}{m}P_{\mathbf{Z}}^i$; therefore, for any function of $\widetilde{\mathbf{Z}}$ such as $f(\widetilde{\mathbf{Z}})$, we have $\E[\ppmZ{m}{1}]{f(\widetilde{\mathbf{Z}})} = \frac{1}{m}\sum_{i=1}^m \E[P_{\mathbf{Z}}^i]{f(\widetilde{\mathbf{Z}})}$. Moreover, if $f$ is invariant with respect to all permutation of elements of $\mathbf{z}$, then $\E[\ppmZ{m}{1}]{f(\widetilde{\mathbf{Z}})} =  \E[P_{\mathbf{Z}}^1]{f(\widetilde{\mathbf{Z}})}$. One can check that this property holds for all moments of  $\widetilde{B}$ and $\widetilde{C}$. Therefore, in the sequel, we assume that $\widetilde{\mathbf{Z}}$ is distributed according to $P_{\mathbf{Z}}^1$. Hence, we have
\begin{align}
\E{\widetilde{B}} &= \E{\frac{1}{m} \sum_{i=1}^m A(\widetilde{Z}_i)} = \frac{\chi_2(Q_1\|Q_0)}{m},\displaybreak[0]\\
\E{\widetilde{B}^2} &= \frac{1}{m^2}\E{\pr{\sum_{i=1}^m A(\widetilde{Z}_i)}^2} = \frac{1}{m^2} \sum_{i=1}^m\sum_{j=1}^m \E{A(\widetilde{Z}_i) A(\widetilde{Z}_j)} = \frac{(m-1)\chi_2(Q_1\|Q_0) + \E[Q_1]{A(Z)^2}}{m^2}\displaybreak[0]\\
&= \frac{\chi_2(Q_1\|Q_0)}{m} + O\pr{\frac{1}{m^2}}\displaybreak[0]\\
\E{\widetilde{B}^3} &= \frac{1}{m^3}\E{\sum_{i=1}^m\sum_{j=1}^m \sum_{k=1}^m A(\widetilde{Z}_i)A(\widetilde{Z}_k) A(\widetilde{Z}_j)} = \frac{\sum_{i=1}^m\E{A(\widetilde{Z}_i)^3} +\sum_{i=2}^m \E{A(\widetilde{Z}_1)A(\widetilde{Z}_i)^2}}{m^3} = O\pr{\frac{1}{m^2}}\displaybreak[0]\\
\E{\widetilde{B}^4} &= \frac{1}{m^4}\E{\sum_{i=1}^m\sum_{j=1}^m\sum_{k=1}^m\sum_{t=1}^m A(\widetilde{Z}_i)A(\widetilde{Z}_j)A(\widetilde{Z}_k)A(\widetilde{Z}_t)}\displaybreak[0]\\
                  &=\frac{\sum_{i=1}^m A(\widetilde{Z}_i)^4 + 2\sum_{i=1}^m \sum_{j=i+1}^m \E{A(\widetilde{Z}_i)^2}\E{A(\widetilde{Z}_j)^2} + \sum_{i=2}^m \E{A(\widetilde{Z}_1)A(\widetilde{Z}_i)^3}}{m^4}\displaybreak[0]\\
&= \frac{\chi_2(Q_1\|Q_0)^2}{m^2} + O\pr{\frac{1}{m^3}}.
\end{align}
One can show that $\abs{\E{\smash{\widetilde{\underline{B}}}^q} - \E{\widetilde{B}^q}} = O(\tau^m)$ as was done earlier for $B$ and $\underline{B}$. Using then the same bounds for $\log(1+x)$ as before and the above calculations, we obtain the desired bounds for the moments of $\widetilde{C}$.
\end{IEEEproof}
}

\bibliographystyle{IEEEtran}

\end{document}